\documentclass[a4paper,11pt]{article} 
\pdfoutput=1 
\usepackage{jheppub} 
\usepackage{amsmath,graphicx,hyperref,chay}

\newcommand{\eps}{\epsilon}
 
\newcommand{\eir}{\epsilon_{\mathrm{IR}}}

\mathchardef\mhyphen="2D
\newcommand{\as}{\alpha_s}

\newcommand{\fms}[1]{{#1}\!\!\!/}

\newcommand{\mc}{\mathcal}
\newcommand{\mr}{\mathrm}

\newcommand{\mO}{\mathcal{O}}

\newcommand{\be}{\begin{equation}} 
\newcommand{\ee}{\end{equation}} 
\newcommand{\bea}{\begin{eqnarray}} 
\newcommand{\eea}{\end{eqnarray}}

\newcommand{\pp}{\perp}
\newcommand{\dg}{\dagger}
\newcommand{\n}{\overline{n}}
 
\newcommand{\nnn}{\frac{\fms{n}}{2}} 

\newcommand{\bl}[1]{{\bf{#1}}}

\newcommand{\blp}[1]{{\bf{#1}}_{\perp}}
\newcommand{\blpu}[1]{{\bf{#1}}^{\perp}}

\newcommand{\bsp}[1]{{\boldsymbol{#1}}_{\perp}}

\newcommand{\nnb}{\nonumber}

\newcommand{\cc}{\mr{c}}
\newcommand{\uc}{\mr{uc}}
\newcommand{\cs}{\mr{cs}}
\newcommand{\ucs}{\mr{ucs}}

\newcommand{\zc}{z_\mr{cut}}

\allowdisplaybreaks[1]

\title{Factorized groomed jet mass distribution\\ in inclusive jet processes}

\def\KU{Department of Physics, Korea University, Seoul 02841, Korea} 
\def\Seoultech{Institute of Convergence Fundamental Studies and School of Liberal Arts, 
Seoul National University of Science and Technology, Seoul 01811, Korea}
\def\Pitt{Pittsburgh Particle Physics Astrophysics and Cosmology Center (PITT PACC) \\ Department of Physics and Astronomy, University of Pittsburgh, Pittsburgh, Pennsylvania 15260, USA}
\author[a]{Junegone Chay}
\emailAdd{chay@korea.ac.kr}
\affiliation[a]{\KU}
\author[b,c]{Chul Kim}
\emailAdd{chul@seoultech.ac.kr}
\affiliation[b]{\Seoultech}
\affiliation[c]{\Pitt}
 
\abstract{ 
We consider the factorized groomed jet mass distribution in inclusive jet processes using modified 
mass drop tagger (mMDT), corresponding to soft drop with the angular exponent $\beta =0$.  
A grooming procedure is implemented rather than tagging in the sense that grooming always returns a groomed jet, while tagging dose not return a jet 
when a single particle remains after tagging.  
We find that the grooming procedure makes the jet mass distribution infrared safe and only ultraviolet divergences appear in each factorized part. 
The groomed jet mass distributions are investigated in a wide range of the jet mass considering various limits on the jet mass variable $\rho = M_J^2/(p_T^JR)^2$ 
and the grooming cut $y_c$. Appropriate effective theories in different kinematic regions are employed to resum large logarithms, in which
the analysis in the region $\rho \sim y_c \ll 1$ is included due to the different type of factorization.
 The analytic computation of the factorized groomed jet mass distribution  is presented 
by resumming the large logarithms in the jet mass, and $y_c$.
Numerically, the effect of the resummation is notably enhanced, compared with the calculation at next-to-leading order, and 
nonglobal logarithms are estimated to be small.
}

\begin{document}

\maketitle 


\baselineskip 3.5 ex 

\section{Introduction}
Study of jet substructure becomes more important in the era of LHC. At high energy, heavy particles such as the $W$, $Z$,
Higgs bosons or the top quark are boosted and the decay products become energetic and collinear, producing jets.  
The signal jets from heavy particles will have different jet substructures compared 
to jets produced by quantum chromodynamics (QCD).  In order to obtain clearer information on the signal jets, it is 
imperative to filter out the background. The background consists of multiple parton scattering and pileup, 
etc., called underlying events,  which will affect the invariant jet mass significantly.
Once we understand how the background can be controlled, we can make more precise description on the signal jet. 
On the other hand, it is also important how the background can be reduced in the jets produced by QCD, which 
also become the background for the signal jets.  Here we consider how to tame the background and
probe the substructure of the QCD jets, which will be the starting point for studying the substructure of the signal jets.

In analyzing jet substructure, there have been many methods which remove wide-angle soft radiation. For example, 
mass drop tagger (MDT) \cite{Butterworth:2008iy} and its modified version (mMDT) \cite{Dasgupta:2013ihk}
were developed and recently soft drop \cite{Larkoski:2014wba} has been suggested, which includes mMDT as a 
special case with the angular exponent $\beta =0$.  Their main goal is taking away soft particles, which are likely to come 
from background, to obtain the information on the hard jets which can be related to the partons participating in the 
hard scattering. 

One of the most important observables in jet substructure is the jet mass distribution
~\cite{Almeida:2008tp,Banfi:2010pa,Li:2011hy,Li:2012bw,Dasgupta:2012hg,Chien:2012ur,Jouttenus:2013hs,Liu:2014oog}, and the plain jet mass distribution 
with small jet radius using the soft-collinear effective theory (SCET) \cite{Bauer:2000ew,Bauer:2000yr,Bauer:2001yt,Bauer:2002nz}
was considered in Refs.~\cite{Chien:2015cka,Hornig:2016ahz,Kolodrubetz:2016dzb,Idilbi:2016hoa,Kang:2018qra}. 
The studies of the groomed jet mass distribution have become rapidly increasing with recent experimental analyses. (See, for example, 
Ref.~\cite{Aaboud:2017qwh}.) 
In Refs.~\cite{Dasgupta:2013ihk,Larkoski:2014wba,Dasgupta:2013via}, the analytical calculation for the jet mass distributions with grooming was extensively 
investigated in QCD. Recently, the groomed jet mass distribution was studied through the resummation of large logarithms 
\cite{Frye:2016okc,Frye:2016aiz,Marzani:2017mva,Marzani:2017kqd,Kang:2018jwa}. The analyses have been performed to next-to-leading logarithmic (NLL) accuracy, 
except  Ref.~\cite{Frye:2016okc,Frye:2016aiz}, in which the groomed jet mass distribution was obtained to next-to-next-to-leading
logarithmic (NNLL) accuracy using soft drop (mMDT as a special case) in SCET.

We consider the groomed jet mass distribution in inclusive jet cross sections, in which an energetic parton can be fragmented to the observed jet. The jet radius $R$ is
supposed to be small, and the jet is described basically by collinear interactions.\footnote{Strictly speaking, our theoretical approach in this paper holds only 
for $R\ll 1$. And the legitimate choice of $R$ for phenomenological study would be $R=0.1-0.4$.
But it has been known that the small $R$ approximation works well even in
the case with $R \gtrsim 0.6$ ~\cite{Dasgupta:2012hg,Jager:2004jh}. So our analysis here can be practically applied to the jet with a sizable value of $R$. }
In this process, we investigate the groomed jet mass distribution from the fragmenting jet 
functions (FJFs)~\cite{Procura:2009vm,Jain:2011xz,Ritzmann:2014mka}. Here the FJFs can provide the detailed information on the jet substructures.
The advantage of considering the groomed jet mass distribution in the FJFs is that the dependence on the jet mass and the grooming parameter resides 
only in this part in the scattering cross section. Since the FJFs describe the properties of the final-state particles, they can be 
studied independent of the scattering processes. That is, we can probe the jet mass distribution in 
hadron-hadron scattering, as well as in $e^+ e^-$ annihilation  with a slight modification of the kinematic variables.

In constructing the jet mass distribution, there is an issue of ``grooming'' versus ``tagging''. Though there is no general agreement on the exact definition on 
grooming and tagging, grooming implies that it always returns an output jet, while tagging sometimes does not~\cite{Dasgupta:2013ihk}. 
Theoretically, this also becomes an issue even at lowest order in which a jet consists of two partons. In tagging, if the energy-cut 
criterion is satisfied, the jet is tagged with the original jet mass as the tagged mass. Otherwise no jet contributes to the
tagged jet mass distribution. The removal of the jet when the tagging criterion is not satisfied renders the jet mass
distribution infrared (IR) sensitive as the jet mass $M_J$ approaches zero. 
In grooming, when the grooming criterion is not satisfied, the final jet consisting of a single energetic parton is included in the jet mass distribution. 
This contribution is concentrated where the jet mass $M_J$ is zero. Therefore adopting grooming means that we add the contribution of 
the $\delta (M_J^2)$ part to the groomed jet mass distribution when the grooming criterion is not satisfied.

In our paper, we choose the groomer in computing the jet mass distributions. In fact,
we show that the groomer is theoretically in a better shape since, after factorization, each factorized part of the groomed jet mass distributions 
turns out to be IR safe. This IR safety enables us to resum large logarithms via 
the renormalization group (RG) equation. We emphasize that the RG equation is meaningful only when there is IR safety, or at least, IR and UV divergences
are separated, so that nonperturbative IR dynamics does not mix with ultraviolet (UV) behavior, as in the case of the parton distribution functions. 
And the normalization of the groomed jet mass 
distributions can be smoothly connected to the plain jet mass distribution without grooming.
Furthermore, after the normalization, the groomed jet mass distribution becomes independent of the renormalization scale.

Introducing the dimensionless variable $\rho = M_J^2/(p_T^J R)^2$, we probe the groomed jet mass distribution 
in a wide range of $\rho$ compared to the grooming parameter $y_c$ in mMDT. 
Here $p_T^J$ is the jet transverse momentum (before grooming)  
relative to the beam axis. In the whole 
range of the jet masses, there are distinct kinematic regions, which show their own characteristics: 
(i) $\rho \sim  y_c\sim \mathcal{O}(1)$.
If we take the limit $y_c\rightarrow 0$, the region corresponds to $y_c\ll \rho \sim \mathcal{O}(1)$, which call  the tail region. 
(ii)  $\rho \ll y_c\sim \mathcal{O}(1)$.
(iii) $\rho \sim y_c\ll 1$, which we call the midrange region. Finally 
(iv) $\rho \ll y_c \ll 1$, which we call the peak region. Note that $y_c$ will be 
fixed at 0.1 in the numerical analysis, but it can be regarded as small or of order 1 compared to the value of $\rho$. 

The tail region corresponds to the ungroomed case, and the resummation of large logarithms of $\rho$ and $y_c~(\zc)$ near the peak region has been considered in 
Refs.~\cite{Kang:2018jwa,Frye:2016okc,Frye:2016aiz}. 
Here we newly include the resummation on the midrange region with $\rho \sim y_c \ll 1$, in which a different factorization structure
is obtained for the resummation of $\ln \rho$ and $\ln y_c$. 
In all these regions, there are various collinear modes with different scaling behavior, which result in different types of
factorization. We employ appropriate effective theories to compute the factorized parts and resum the large logarithms of $\ln \rho$ or $\ln y_c$. 
Also we confirm that all the factorized functions are IR safe and contain only UV divergence when the grooming procedure is applied.
Based on the results, we are able to describe the behavior of the groomed jet mass 
distribution in a wide range of $\rho$, say, from $\mc{O}(10^{-5})$ to $\mc{O}(1)$, with different forms of factorization. 
 
The organization of the paper is as follows: In Sec.~\ref{ungjff}, the plain jet mass distribution from the FJFs before grooming is briefly described, and it offers the basis for defining the groomed jet mass distribution. 
In Sec.~\ref{grjff}, we explain how the grooming procedure is applied to the FJFs, and define the groomed jet mass
distribution. The theoretical implementation of grooming is described in detail, especially about how to obtain the IR safe results.
In Sec.~\ref{grjm}, we identify important modes in different regions, employ the relevant effective theories, and compute
the factorized parts of the groomed jet mass distribution to next-to-leading order (NLO) in $\as$. 
In Sec.~\ref{rsgjm}, we resum the large logarithms that appear in the midrange and the peak regions to NLL accuracy. 
In Sec.~\ref{num}, numerical analysis is performed for the groomed jet 
mass distributions in the whole range of the jet mass by interpolating the resummed results in various regions. The effect of the 
nonglobal logarithms is also estimated. In Sec.~\ref{conc}, we present conclusions. 

In Appendix~\ref{apa}, we list all the functions appearing in the text. In Appendix~\ref{apb}, the 
relation between the $\Lambda$-distribution and the standard plus  distribution functions is given. 
In Appendix~\ref{apc}, details in obtaining $\tilde{\mathcal{C}}_k (y_c, Q^2, \mu)$ in Case (i), and 
$\mathcal{S}_k^{II} (y_c^2 Q^2, \mu)$ in Case (iv) are presented with the structure of the phase spaces. 
In Appendix~\ref{apd}, we show the results in the midrange and the peak regions, employing soft drop with $\beta \ge 0$.

\section{Jet mass distribution from jet fragmentation function\label{ungjff}}

We consider the inclusive jet cross section in $N_1 N_2 \rightarrow JX$, where $N_1$, $N_2$ are incoming hadrons, 
$J$ is the jet with small radius $R$, and $X$ denotes all the remaining particles. The scattering cross section is written
as~\cite{Dasgupta:2014yra}
\begin{equation}
\label{jsec}
\frac{d\sigma}{dy dp_T^J} =\sum_i \int_{x_J=p_T^J/Q_T}^1 \frac{dx}{x} 
\frac{d\sigma_i (x_J/x,\mu)}{dy dp_T^i} D_{J/i} (x;p_T^i R,\mu),
\end{equation}
where $\sigma_i$ is the partonic cross section in which the final-state jet is produced by the parton $i$, with the 
maximum transverse momentum $Q_T$ with respect to the beam direction for a given rapidity $y$.  The function 
$D_{J/i}$, which we call  the fragmentation function to a jet (FFJ), describes the probability of  producing the outgoing jet 
$J$ from the mother parton $i$ with the momentum fraction $x$. The jet cross section in Eq.~(\ref{jsec}) necessarily 
involves large logarithms of small $R$, which can be resummed through the DGLAP evolutions of the
FFJs~\cite{Dasgupta:2014yra,Dasgupta:2016bnd,Kaufmann:2015hma,Kang:2016mcy,Dai:2016hzf}. 

Throughout this paper, we use the $\mr k_{\mr{T}}$-type algorithms in defining the jet, which include 
the $\mr k_{\mr{T}}$~\cite{Catani:1993hr,Ellis:1993tq},  the anti-$\mr k_{\mr{T}}$~\cite{Cacciari:2008gp}, and the 
Cambridge/Aachen (C/A) algorithms~\cite{Dokshitzer:1997in}. 
Up to NLO in $\as$ (two particles in a jet at most), these three algorithms have the same constraint for merging into a jet with small 
$R$ as  
\be \label{merging}
\theta < R'~ 
\left\{
\begin{array}{rl} 
& R'= R ~~\mbox{for $e^+e^-$ annihilation} \\ 
&R'=R/\cosh y  ~~\mbox{for hadron collision.} \end{array} \right.
\ee
Then the typical jet scale $E_JR'$ is written as $p_T^JR$ for hadron-hadron collisions and $E_JR$ for $e^+e^-$ annihilation. 

We go further to consider the fragmentation inside the
jet~\cite{Procura:2011aq,Kaufmann:2015hma,Dai:2016hzf,Kang:2016ehg}. 
To NLO in $\as$, the factorized cross section reads~\cite{Dai:2016hzf}
\begin{equation} \label{defjff}
\frac{d\sigma}{dy dp_T^J dz} = \sum_{i,k} \int_{x_J}^1 \frac{dx}{x} \frac{d\sigma_i (x_J/x, \mu)}{dy dp_T^i} 
D_{J_k /i}(x;p_T^i R,\mu) D_{l/J_k} (z;p_T^JR). 
\end{equation}
Here $l$ is the hadron or the subjet inside the jet $J_k$ initiated by the parton $k$, and $z= p_T^l /p_T^J$ is the 
momentum fraction of $l$ with respect to $J_k$.  We call $D_{l/J_k} (z)$ the jet fragmentation function (JFF) 
for $l$, which is the probability for the fragmentation process $k \rightarrow l$ inside the jet $J_k$.  When we consider 
jet substructure, the advantage in focusing on the JFF is because it is process-independent and is also
independent of the renormalization scale. The remaining product of  the partonic cross section with the FFJ
in Eq.~(\ref{defjff}) is scale invariant as well~\cite{Dai:2016hzf}. 

The FFJ and the JFF are the probabilities, satisfying the momentum sum rules  
\begin{equation}
\sum_k \int_0^1 dx x D_{J_k/i} (x; p_T^i R,\mu) =1, \ \ \sum_l \int_0^1 dz z D_{l/J_k} (z; p_T^J R) =1. 
\end{equation}
The fragmenting processes from the quark  and the gluon jets in SCET are described by  
\begin{equation}
\begin{split}
\tilde{D}_{l/J_q} (z,\mu)  &=  \sum_{X\in J} \frac{1}{2N_c z} \int d^{D-2} \mathbf{p}_{\perp}  \\
 &\times \mathrm{Tr} \langle 0| \delta \Bigl(
\frac{p_+}{z} -\mathcal{P}_+ \Bigr) \delta^{(D-2)} (\bsp{\mc{P}}) 
\frac{\FMslash{\overline{n}}}{2}\chi_n |l(p_+, \mathbf{p}_{\perp}) X\rangle  
\langle l(p_+, \mathbf{p}_{\perp})  X| \overline{\chi}_n |0\rangle,   \\
\tilde{D}_{l/J_g} (z,\mu)  &=  \sum_{X\in J} \frac{1}{z p_J^+ (D-2) (N_c^2 -1)} \int d^{D-2} \mathbf{p}_{\perp}   \\
 &\times \mathrm{Tr} \langle 0| \delta \Bigl(
\frac{p_+}{z} -\mathcal{P}_+ \Bigr) \delta^{(D-2)} (\bsp{\mc{P}})  \mathcal{B}_n^{\perp \mu,a}
|l(p_+, \mathbf{p}_{\perp}) X\rangle \langle l(p_+, \mathbf{p}_{\perp})  X| \mathcal{B}_{n,\mu}^{\perp ,a} |0\rangle,
\end{split}
\end{equation}
where the spacetime dimension is $D= 4-2 \eps$, and $\chi_n = W_n^{\dagger} \xi_n$ is the collinear quark field in the 
$n$-lightcone  direction, with the collinear Wilson line
$W_n$. And $\mc{B}_{n}^{\pp\mu,a}  = i\n^{\rho}g_{\perp}^{\mu\nu} G_{n,\rho\nu}^b \mc{W}_n^{ba} 
= i\n^{\rho}g_{\perp}^{\mu\nu} \mc{W}_n^{\dagger,ba} G_{n,\rho\nu}^b$ is the collinear gluon field strength tensor, 
where $\mc{W}_n$ is the Wilson line in the adjoint representation. In our convention 
$p_+\equiv \n\cdot p = p_0 + \hat{\bl{n}}_J\cdot \bl{p}$, $p_-\equiv n\cdot p = p_0 -\hat{\bl{n}}_J\cdot \bl{p}$, 
where $n^2 = \overline{n}^2=0$, $n\cdot \overline{n} =2$, and $\hat{\bl{n}}_J$ is the unit vector along the jet direction.

Note that $\tilde{D}_{l/J_k}$ and the integrated jet function $\mathcal{J}_k$ within the jet are related by
\begin{equation}
\sum_l \int_0^1 dz z \tilde{D}_{l/J_k} (z; p_T^J R,\mu) =\mathcal{J}_k (p_T^J R,\mu).
\end{equation}
To one loop, the integrated jet functions with the $\mr k_{\mr{T}}$-type algorithms 
are given as ~\cite{Procura:2011aq,Cheung:2009sg,Ellis:2010rwa,Chay:2015ila}
\bea\label{qintj} 
\mc{J}_q(p_T^JR,\mu) &=& 1+\frac{\as C_F}{2\pi} \Biggl[
\frac{3}{2}\ln\frac{\mu^2}{p_T^{J2}R^2}+\frac{1}{2}\ln^2\frac{\mu^2}{p_T^{J2}R^2}+\frac{13}{2}-\frac{3\pi^2}{4} \Biggr]\ , 
\\
\mc{J}_g(p_T^JR,\mu) &=& 1+\frac{\as C_A}{2\pi} \Biggl[
\frac{\beta_0}{2C_A}\ln\frac{\mu^2}{p_T^{J2}R^2} 
\label{gintj}
+\frac{1}{2}\ln^2\frac{\mu^2}{p_T^{J2}R^2}+\frac{67}{9}-\frac{23n_f}{18C_A}-\frac{3\pi^2}{4} \Biggr]\ ,
\eea
where $\beta_0 = 11N_c/3-2n_f/3$, $C_A=N_c=3$, and $n_f$ is the number of quark flavors.
The JFF $D_{h/J_k}$ is normalized from $\tilde{D}_{h/J_k}$ divided by $\mathcal{J}_k $, and is given by 
\begin{equation}
\label{normjff}
D_{l/J_k} (z; p_T^J R) = \frac{\tilde{D}_{h/J_k} (z; p_T^J R,\mu)}{\mathcal{J}_k (p_T^J R,\mu)}\ .
\end{equation}
Since $\mc{J}_k$ is included in the FFJs $D_{J_k /i}(x)$ in Eq.~(\ref{defjff})~\cite{Dai:2016hzf}, there is no double counting 
of $\mc{J}_k$.  

In order to consider the jet mass distribution for the inclusive jet production,
we focus on the jet mass distribution of the JFF, i.e., the fragmenting jet function
(FJF)~\cite{Procura:2009vm,Jain:2011xz,Ritzmann:2014mka}. The normalized FJF within the jet, $\mc{G}_{l/J_k}$, is defined as
\begin{equation}
\label{FJF}
D_{l/J_k} (z) =\int dM_J^2~\mc{G}_{l/J_k} (z,M_J^2). 
\end{equation}
Applying the momentum sum rule,  the plain jet mass distribution is given by~\cite{Idilbi:2016hoa} 
\be
\label{sumFJF}
\Phi_k^{\mr{pl}}(M_J^2;p_T^J R) = \sum_{l=q,\bar{q},g} \int^1_0 dz z~ \mc{G}_{l/J_k} (z,M_J^2).
\ee 
The plain jet mass distribution is scale invariant, and normalized to one since it satisfies
\be
\int dM_J^2 \, \Phi_k^{\mr{pl}}(M_J^2;p_T^J R) =  
\sum_{l=q,\bar{q},g} \int^1_0 dz z~D_{l/J_k} (z)= 1. 
\ee

Applying the momentum sum rule for $z=p_T^l/p_T^J$ in Eq.~(\ref{defjff}), 
the factorization theorem for the cross section with the jet mass $M_J$ and the transverse jet momentum 
$p_T^J$ is written as 
\begin{equation}
\label{factjm}
\frac{d\sigma}{ dy dp_T^J dM_J^2} =\sum_{i,k} \int_{x_J}^1 \frac{dx}{x} \frac{d\sigma_i (x_J/x, \mu)}{dy dp_T^i} 
D_{J_k /i}(x; p_T^i R,\mu) \Phi_k^{\mr{pl}}(M_J^2;p_T^J R). 
\end{equation}
Eq.~(\ref{factjm}) is our starting point to study the groomed jet mass distribution for the inclusive jet process.
From Eq.~\eqref{sumFJF}, the grooming condition in mMDT can be accomplished by restricting the range of 
$z$ to $y_c/(1+y_c) < z < 1/(1 +y_c)$. 
In Eq.~(\ref{factjm}), if $x$ is not too close to one and $M_J \sim p_T^JR$, the FFJs $D_{J_k/i}$ and the jet mass 
distributions $\Phi_k^{\mr{pl}}$ can be described solely by the collinear modes scaling as 
$(p_+,p_-,\blp{p})\sim p_T^J(1,R^2,R)$. 

If the observables are sensitive to the 
collinear-soft (csoft) radiations~\cite{Bauer:2011uc,Procura:2014cba,Becher:2015hka,Chien:2015cka}, 
$\Phi_k^{\mr{pl}}$ and $D_{J_k/i}$ can be refactorized by including the csoft 
interactions~\cite{Idilbi:2016hoa,Kang:2018jwa,Dai:2017dpc,Liu:2017pbb}. 
For example, when $M_J \ll p_T^J R$, $\Phi_k^{\mr{pl}}(M_J^2)$ is  refactorized as~\cite{Idilbi:2016hoa}
\be
\label{facpljm}
\Phi_{k}^{\mr{pl}}(M_J^2\ll p_T^{J2}R^2;p_T^JR) = \mc{C}_k(p_T^{J2}R^2,\mu)\int_0^{M_J^2} dM^2 J_{k} (M^2;\mu) 
{S}_{k} (M_J^2-M^2;p_T^JR,\mu), 
\ee
where the collinear functions $\mc{C}_k$ are given by the inverse of the integrated jet functions $\mc{J}_k^{-1}$ for $k=q, g$, and enter as
the normalization. $J_k (M^2)$ are the ``standard jet functions'' introduced in
Refs.~\cite{Korchemsky:1994jb,Akhoury:1995fp,Bauer:2001yt,Bosch:2004th}, where $J_q$ is defined as 
\be
\sum_{X_n} \langle 0 | \chi_n^{\alpha} |X_n \rangle \langle X_n | \bar{\chi}_n^{\beta} | 0 \rangle 
\label{jetfq}
=\int \frac{d^4 p}{(2\pi)^3} p_+ \nnn J_{q}(p^2,\mu) \delta^{\alpha\beta},
\ee
and $J_g$ is similarly defined in terms of $\mc{B}_{n}^{\pp\mu,a}$. The functions $J_k$ are governed by 
the ``ultracollinear'' modes, which scale as $p_{\mr{uc}} = (p_{\uc}^+,p_{\uc}^-,\blpu{p}_{\uc}) 
= p_T^J(1,M_J^2/p_T^{J2},M_J/p_T^J)$.

Finally the csoft functions $S_k$ describe the interaction of the csoft modes, which scale as 
\be\label{spcs}
p_{\cs}^{\mu} = (p_{\cs}^+,p_{\cs}^-,\blpu{p}_{\cs}) \sim \frac{M_J^2}{p_T^J R^2} (1,R^2,R) = \rho p_T^J (1,R^2,R).
\ee 
The quark csoft function $S_q$ is defined as 
\be
\label{sqf}
S_q(M^2=p_J^+\ell_-,\mu) =  \frac{1}{p_J^+ N_c} \mr{Tr}~\langle 0 | Y_{\n,\cs}^{\dg} Y_{n,\cs}
\delta(\ell_-+\Theta(R-\theta) i\partial_- ) Y_{n,\cs}^{\dg} Y_{\n,\cs} |0\rangle_, 
\ee
where $Y_{n(\n),\cs}$ are the csoft Wilson lines, in which  the soft  gauge field in the soft Wilson lines~\cite{Bauer:2001yt} 
is replaced by the csoft gauge field.  
The gluon csoft function $S_g$ is expressed in terms of the csoft Wilson lines in the adjoint representation.
Note that the collinear and csoft modes are sensitive to the jet boundary characterized by the jet radius $R$, 
while the ultracollinear modes are too narrow to recognize it. The large energy hierarchy between the collinear and csoft 
modes could give rise to large nonglobal logarithms (NGLs)~\cite{Dasgupta:2001sh,Banfi:2002hw}, which give sizable uncertainty 
in estimating the small jet mass distributions. 
 
\section{Jet grooming and the groomed jet mass distribution\label{grjff}}
We employ mMDT~\cite{Dasgupta:2013ihk} for grooming, which involves two parameters $y_c$ and $\mu$.
It is prescribed as follows:  For an initial jet $j$,
\begin{enumerate}
\item Decluster the jet $j$ into two subjets $j_1$ and $j_2$ with $m_{j_1} > m_{j_2}$ by undoing the last clustering process.
\item If there is a significant mass drop, $m_{j_1} < \mu m_j$, and the splitting satisfies the criterion
\be\label{mdtc}
\mathrm{min} (p_{Tj_1}^2, p_{Tj_2}^2) 
\frac{\Delta R_{j_1 j_2}^2}{m_j^2} > y_c, 
\ee
then take $j$ to be the tagged jet.
\item Otherwise redefine $j$ to be that of $j_1$ and $j_2$ with the larger transverse mass $m^2 + p_T^2$ and go back to 
step 1 (unless $j$ consists of a single particle, in which case the original jet is deemed untagged.)
\end{enumerate}

At leading order in which the jet consists of two partons, when the jet is declustered, each jet is massless 
and the mass drop condition is automatically satisfied and the parameter $\mu$ is irrelevant. Then, 
if the criterion in Eq.~\eqref{mdtc} is satisfied, the jet is tagged. Otherwise, the jet does not contribute to the tagged
jet mass distribution.  On the other hand, soft drop has a more generalized criterion. Compared
to the mMDT condition in Eq.~\eqref{mdtc}, it is given as
\begin{equation} \label{sd}
\frac{\mathrm{min} (p_{t1}, p_{t2})}{p_{t1}+p_{t2}} > z_{\mathrm{cut}} \Bigl(\frac{\theta_{12}}{R}\Bigr)^{\beta},
\end{equation}
where $z_{\mathrm{cut}}$ has the same role as $y_c$ in mMDT and $\beta$ is the angular exponent 
to control the dependence of the angle between the two partons,
$\theta_{12} = \sqrt{\Delta y_{12}^2 +\Delta\phi_{12}^2}$. The limit $\beta \rightarrow 0$ corresponds to mMDT. 
We will also use soft drop in the midrange and peak regions.

This is the original prescription, but we further implement the idea of ``grooming'', 
which always returns a groomed jet. 
It means that, when a single particle remains in the jet after grooming, 
we include the contribution of the single particle to the groomed jet mass distribution. That is, we include the contribution of the 
jet mass with $M_J =0$.  Even though the criterion for mMDT or soft drop is 
not satisfied, the remaining particles contribute to the jet mass. It also holds when only a single particle remains in the jet. 
Therefore the ``groomed jet mass distribution'' starts from $\delta(M_J^2 )$ at order $\as^0$. 
Also at NLO, we include the virtual corrections to the jet mass distribution, which cancel the IR divergence in the real emissions. 
This makes the resultant groomed jet mass distribution IR safe even in the limit $M_J \to 0$. 
Since we do not drop any event as in the plain jet mass, the normalization is the same as in the plain jet 
mass distribution, i.e., 
\be 
\label{jmnorm}
\int dM_J^2~\Phi_{k=q,g} (y_c,M_J^2) = \int dM_J^2~\Phi_{k}^{\mr{pl}} (M_J^2)= 1,
\ee
where $\Phi_k(y_c,M_J^2)$ is the groomed jet mass distribution. Despite the same normalization, 
the dependence of the groomed jet mass distribution on $M_J$ is different from that of the plain jet mass distribution since the jet mass is different
in the regions where the criterion for mMDT in Eq.~(\ref{mdtc}), or for soft drop 
in Eq.~(\ref{sd}) is not satisfied.
 
By applying grooming with mMDT to Eq.~(\ref{sumFJF}), as prescribed above, the groomed jet mass distribution at NLO in $\as$ can 
be expressed as
\bea
\label{gjmexp}
\Phi_k (y_c, M_J^2)  &=&  \sum_{l=q,\bar{q},g} \Biggl[\int^{z_{\mathrm{max}}}_{z_{\mathrm{min}}} dz 
z~\mc{G}_{l/J_k} (z,M_J^2) \nnb \\
&&+\delta (M_J^2) \Bigl(\int_0^{z_{\mathrm{min}}} dz +\int_{z_{\mathrm{max}}}^1 dz \Bigr) 
\int^{M^2_{\mathrm{max}}}_0 dM^2 z~\mc{G}_{l/J_k} (z,M^2)\Biggr]_,
\eea
where $z_{\mathrm{min}} = y_c/(1+y_c)$ and $z_{\mathrm{max}} = 1/(1+y_c)$, and 
$M_{\mathrm{max}}^2 = z(1-z) p_T^{J2}R^2$. At NLO, if the criterion Eq.~(\ref{mdtc}) is not satisfied, only a single 
energetic particle is included. It is given by the second term in Eq.~(\ref{gjmexp}), proportional to $\delta(M_J^2)$, 
and the virtual corrections are included. 

If the second term is discarded in Eq.~(\ref{gjmexp}), it corresponds to the jet tagging, in  which no single energetic parton 
is included in the final jet. We can show that the jet mass distribution in this case becomes IR sensitive as $M_J$ goes to 
zero from explicit calculations at NLO. Let us consider the region $\rho  \sim y_c \sim \mc{O}(1)$ as an example. In this region, only 
the collinear modes with $(p_c^+,p_c^-,\blpu{p}_c)\sim p_T^J(1,R^2,R)$ contribute to the jet mass distribution.  We obtain 
the ``tagged jet mass distribution'' as
\begin{eqnarray}
\Phi_{k=q,g}^{\mathrm{tag}} (y_c, M_J^2) &=& \frac{\as C_k}{2\pi}\Biggl\{-\delta (M_J^2) \Bigl(  \frac{1}{\eir} 
+\ln \frac{\mu^2}{M_c^2} + h_k (y_c) \Bigr)  \nonumber \\
\label{tagjm}
&&
+ f_k (w) \frac{\theta (M_J^2 -M_c^2)}{M_J^2} + g_k (y_c) \Bigl[\frac{\theta(M_c^2 -M_J^2)}{M_J^2}\Bigr]_{M_c^2}\Biggr\},  
\end{eqnarray}
where $C_k= C_F, C_A$ for $k=q, g$, $M_c^2 = p_T^{J2} R^2 y_c/(1+y_c)^2$ and $w = \sqrt{1-4M_J^2/(p_T^{J} R)^2}$.  
The functions $f_k (y)$, $g_k (w)$, and $h_k (y)$ are given in Eq.~(\ref{fqg}), Eq.~(\ref{gqg}), and Eq.~(\ref{hqg}) respectively. 
In extracting the  IR divergence as $M_J^2$ goes to zero, we employ the ``$\Lambda$-distribution'', which is defined as
\begin{equation}
\label{lamdad}
\int_0^{\mathcal{M}^2} dM^2 [G (M^2)]_{\Lambda^2} F(M^2) = \int_0^{\mathcal{M}^2} dM^2 G(M^2) F(M^2) 
-\int_0^{\Lambda^2} dM^2 G(M^2) F(0),
\end{equation}
for any smooth function $F(M^2)$ near $M^2 =0$. In Eq.~(\ref{tagjm}), $\Lambda^2$ has been set to $M_c^2$.

Due to the IR divergence proportional to $\delta(M_J^2)$ in Eq.~(\ref{tagjm}), it is not possible to normalize the 
distribution unless we impose a small nonzero jet mass cut, $M_{J,\mathrm{cut}}$. Then the normalization 
of the tagged jet mass distribution depends on the mass cut, which results in unwanted uncertainty. For example, if 
$M_{J,\mathrm{cut}} \ll p_T^JR$, the normalization involves a large logarithm involving the mass cut.
 
On the other hand, the groomed jet mass distribution is IR safe since the second term in Eq.~(\ref{gjmexp}) cancels the 
IR divergence in the tagged distribution. The NLO result reads 
\bea \label{phi1}
\Phi_k (y_c, M_J^2) &=& \delta (M_J^2) \Bigl(1-\frac{\alpha_s C_k}{2\pi} I_k (y_c)\Bigr) \nnb \\
&&+\frac{\alpha_s C_k}{2\pi} \Biggl\{
 f_k (w) \frac{\theta (M_J^2 -M_c^2)}{M_J^2} + g_k (y_c) \Bigl[\frac{\theta(M_c^2 -M_J^2)}{M_J^2}\Bigr]_{M_c^2}\Biggr\}, 
\eea
where the functions $I_k (y)$ are given in Eq.~(\ref{iqg}). Note that the groomed jet mass distribution 
in Eq.~(\ref{phi1}) is the same as the tagged case in Eq.~(\ref{tagjm}) except the terms proportional to $\delta (M_J^2)$. 
Therefore, for the jet mass distribution, the distinction between ``jet grooming'' and ``jet tagging'' is related to the issue on 
how to treat the jet with $M_J =0$, e. g., with a single parton in the jet. 

There is a theoretical advantage in taking grooming over tagging. 
First, we can properly normalize the jet mass distribution in Eq.~(\ref{jmnorm}) without imposing the mass cut.  
Second, since we keep the leading-order (LO) result, the nonzero $M_J$ part appears at NLO in the grooming
procedure.\footnote{\label{tag}In tagging,  since the second term in Eq.~(\ref{gjmexp}) is removed, 
the LO result appears at order $\as$, hence the NLO result is counted as $\mc{O}(\as^2)$. Therefore two-loop calculations 
for the jet mass are needed in order to resum the large logarithms via the RG equations in the tagging
procedure. } This makes it convenient to factorize the jet mass distribution when we consider various small limits of 
$\rho$ and $y_c$. Through the NLO calculations in various limits, we can understand the factorization structure and 
its consistency transparently. Moreover, the resummation of the large logarithms arising from small $\rho$ and $y_c$ 
can be systematically performed mainly from the NLO results of each factorized part, as we will show later. 
In the next section, we investigate the groomed jet mass distributions in various limits of $\rho$ and $y_c$. We employ
appropriate effective theories for different modes in each region to establish the factorization. 

\section{Groomed jet mass distribution in various regions\label{grjm}}
In Eq.~(\ref{phi1}), we have considered the groomed jet mass distribution in the tail region $\rho \sim y_c \sim \mc{O}(1)$ 
without taking specific limits on $\rho= M^2/(p_T^{J}R)^2$ and $y_c$.
In this case,  we can describe the distribution only with the collinear modes scaling as 
$(p_c^+,p_c^-,\blpu{p}_c)\sim p_T^J(1,R^2,R)$. In this section, we consider the cases with small jet mass for different
hierarchies of $\rho$ and $y_c$. They are given as
\begin{equation}
\label{vlimit}
(a) \  \rho \ll  y_c \sim \mathcal{O}(1), \ ~
(b) \ y_c \ll 1, ~
\left\{
\begin{array}{rl} 
& \rho \sim y_c: ~~\mbox{the midrange region} \\ 
& \rho \ll y_c: ~~\mbox{the peak region.} \end{array} \right.
\end{equation}
Since the jet mass is small,  there exist characteristic scales smaller than the collinear scale $\mu_{\mr{c}} \sim p_T^JR$. In fact,
various subsets of the collinear modes are required to describe the dynamics at lower scales than the collinear scale.    
  
\begin{figure*}[t]
\begin{center}
\includegraphics[width=16cm]{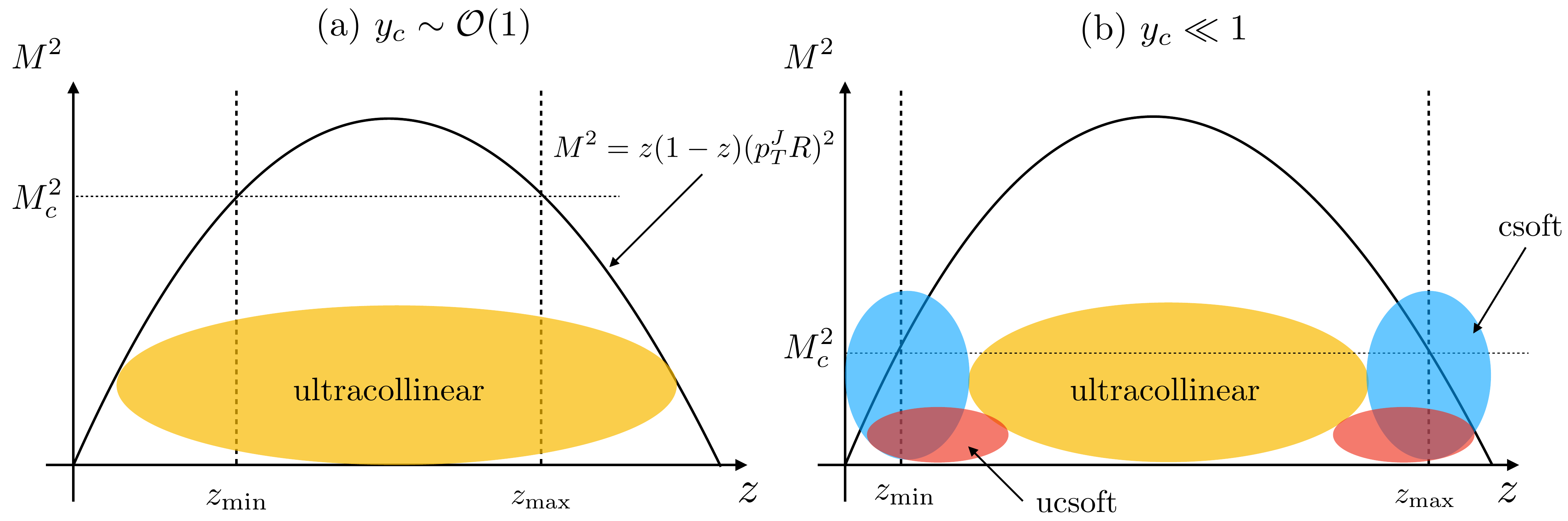}
\end{center}
\vspace{-0.6cm}
\caption{\label{fig-1} \baselineskip 3.0ex 
Subsets of the collinear modes are shown in the $z$-$M^2$ space for (a) $y_c \sim \mc{O}(1)$ and (b) $y_c \ll 1$. 
Here $z$ is the energy fraction of a parton and $M^2$ is the invariant mass squared. 
The parabola is the jet boundary, dictated by the $\mr{k_T}$-type algorithm. The region between $z_{\mr{max}}$ and
$z_{\mr{min}}$ under the parabola is the region satisfying the mMDT criterion,  Eq.~(\ref{mdtc}).
} 
\end{figure*}

In case (a) with $\rho \ll y_c \sim \mathcal{O}(1)$, the collinear mode cannot contribute to the small jet mass since 
$p_c^2 \gg M_J^2$. Instead the ultracollinear modes contribute to the jet mass, with the scaling 
\begin{equation}
\label{ucscale}
p_{\mathrm{uc}}^{\mu} \sim  p_T^J \Bigl(1, \frac{M_J^2}{p_T^{J2}}, \frac{M_J}{p_T^J}\Bigr) = p_T^J (1, \rho R^2, \sqrt{\rho} R),
\end{equation}
where $p_{\mathrm{uc}}^2 \sim M_J^2 = \rho (p_T^J R)^2$, which is suppressed by $\rho$ compared to $p_c^2$. 
The ultracollinear mode is insensitive to the jet boundary.  Fig.~\ref{fig-1}-(a) shows the ultracollinear mode for case (a) in Eq.~\eqref{vlimit}, responsible for narrow
energetic radiations inside the jet. Since $z_{\mr{max}}$ $(z_{\mr{min}})$ is of $\mc{O}(1)$, csoft and soft radiations do not contribute.
  
In case (b) with $y_c \ll 1$ in Eq.~\eqref{vlimit}, in addition to the ultracollinear mode, there exist other modes due to the smallness of $y_c$. In the 
midrange region where $\rho \sim y_c~ (M_J^2 \sim M_c^2 \sim y_c (p_T^J R)^2)$, the csoft mode can contribute to the
nonzero groomed jet mass in addition to the ultracollinear mode, as illustrated in Fig.~\ref{fig-1}-(b). 
Here the csoft momentum scales as 
\be\label{csofts}
p_{\mathrm{cs}}^{\mu} \sim y_c p_T^J (1, R^2, R)~~\text{with~} 
p_{\mathrm{cs}}^2 \sim (y_c p_T^J R)^2.
\ee 
Hence the csoft contribution to the jet mass 
$M_{\cs}^2 = p_J^+ p_{\cs}^-$ is power counted as $y_c (p_T^J R)^2$ and comparable to the jet mass squared in this region. 
The csoft mode is the scaled-down version of the collinear mode, hence sensitive to the jet boundary, as shown 
in Fig.~\ref{fig-1}-(b).
 
In the peak region with $\rho \ll y_c \ll 1$, we need much narrower version of the csoft mode to produce much smaller
nonzero jet mass. We call this mode the ultracollinear-soft (ucsoft) mode, which scales as
\be
\label{pucs} 
p_{\mathrm{ucs}}^{\mu} \sim y_c p_T^J \Bigl(1,  \frac{M_J^2}{y_c p_T^{J2}}, \frac{M_J}{\sqrt{y_c}p_T^{J}}\Bigr) 
\sim y_c p_T^J \Bigl(1,  R^2\frac{\rho}{y_c},  R\sqrt{\frac{\rho}{y_c}} \Bigr),  
\ee
with $p_{\mathrm{ucs}}^2 \sim y_c M_J^2$. Therefore the ucsoft mode is much narrower than the csoft mode 
although the largest momentum component is of the same magnitude $y_c p_T^J$. The ucsoft mode cannot read the jet
boundary, as can be seen in Fig.~\ref{fig-1}-(b).
 
\subsection{$\rho \ll y_c \sim \mathcal{O} (1)$}
\begin{figure*}[t]
\begin{center}
\includegraphics[width=16cm]{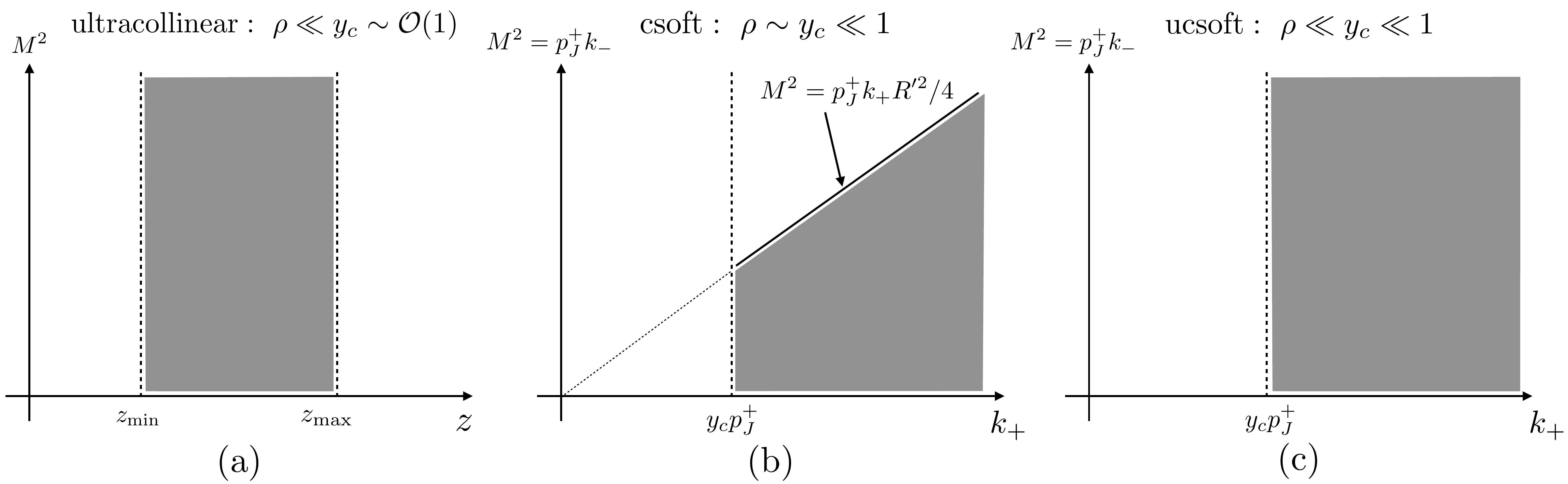}
\end{center}
\vspace{-0.6cm}
\caption{\label{fig0} \baselineskip 3.0ex 
Phase spaces (shaded regions) for the submodes satisfying the mMDT criterion in Eq.(\ref{mdtc}) at one loop. 
Diagrams (b) and (c) denote the phase spaces for the csoft and ucsoft modes with the momentum $k$. 
The nonzero groomed mass comes from the shaded regions, and $\delta(M_J^2)$ comes from the remaining regions 
according to grooming.
} 
\end{figure*}

In this case, the ultracollinear mode entirely describes the groomed nonzero jet mass, but this mode is insensitive to the jet boundary. 
In Fig.~\ref{fig0}-(a), the phase space for the ultracollinear mode is shown. The shaded region satisfies the mMDT criterion in Eq.~\eqref{mdtc} and contributes to the 
 nonzero groomed mass. Since the invariant mass squared for the ultracollinear mode is small, the upper bound for $M^2$ extends to infinity.
Outside the shaded region, $\delta(M_J^2)$ is returned since the remaining single energetic parton is included  
in the ``grooming''  procedure. As a result, at NLO in $\as$, the groomed jet functions from the ultracollinear mode are expressed
as ($k=q$, $g$)
\bea
\tilde{J}_k (M_J^2,\mu)  &=&  \sum_{l=q,\bar{q},g} \Biggl[\int^{z_{\mathrm{max}}}_{z_{\mathrm{min}}} 
dz z~\tilde{\mc{G}}_{l/k} (z,M_J^2) \nnb 
\\
&&+\delta (M_J^2) \Bigl(\int_0^{z_{\mathrm{min}}} dz +\int_{z_{\mathrm{max}}}^1 dz \Bigr) 
\int^{\infty}_0 dM^2 z~\tilde{\mc{G}}_{l/k} (z,M^2)\Biggr]  \nonumber\\
\label{gjf}
&=& \delta (M_J^2) +\frac{\alpha_s C_k}{2\pi} \Bigl\{ -\delta (M_J^2) 
\Bigl[  g_k (y_c)\ln \frac{\mu^2}{\Lambda^2}    + h_k (y_c)\Bigr] + g_k (y_c) \Bigl[ \frac{1}{M_J^2}\Bigr]_{\Lambda^2} \Bigr\}.
\eea  
Here $\tilde{\mc{G}}_{l/k} (z,M^2)$ is the generic FJFs introduced in Ref.~\cite{Procura:2009vm,Jain:2011xz}.   
Unlike $\mc{G}_{l/J_k}$ that was introduced in Eq.~(\ref{FJF}), $\tilde{\mc{G}}_{l/k}$ is not divided by $\mc{J}_k$. 
The functions $g_k$ and $h_k$  in Eq.~(\ref{gjf}) are given in Eqs.~(\ref{gqg}) and (\ref{hqg}) respectively. 

Even though the collinear mode does not contribute to the groomed nonzero jet mass, it can radiate in the 
regions $[0,z_{\mr{min}})$ and $(z_{\mr{max}},1]$ inside a jet.
Employing the zero-bin subtraction \cite{Manohar:2006nz} to avoid double counting on the phase space overlapped 
with the ultracollinear mode, the one-loop result for the collinear mode is IR finite and the divergence is of the UV
origin. The renormalized collinear contribution to NLO in $\as$ is given by 
\begin{equation} \label{hard2}
\tilde{\mc{C}}_k (Q^2, \mu) = 1+\frac{\as C_k}{2\pi} \Bigl[ g_k (y_c)  \ln \frac{\mu^2}{M_c^2}  +h_k (y_c) -I_k (y_c)\Bigr],
\end{equation}
where we use $Q \equiv p_T^J R$ for simplicity, and $M_c^2 = Q^2 y_c/(1+y_c)^2$. In Appendix~\ref{apc} we show the details of the calculation.  

As a result, the groomed jet mass distribution functions $\Phi_k$ ($k=q$, $g$) in the limit 
$\rho \ll y_c \sim \mc{O}(1)$ are factorized as
\begin{equation}
\label{srho}
\Phi_k (y_c\sim \mO(1), M_J^2\ll Q^2) = \tilde{\mc{C}}_k (Q^2, \mu) \tilde{J}_k (M_J^2,\mu). 
\end{equation}
Combining Eqs.~(\ref{gjf}) and (\ref{hard2}), it can be verified that the NLO result in Eq.~(\ref{srho}) reproduces 
the full result in Eq.~(\ref{phi1}) in the limit $M_J^2 \ll Q^2$.
Note that each factorized function in Eq.~(\ref{srho}) contains a single logarithm at one loop. It implies that the dominant
logarithmic corrections appear as $\sum_{n=0} c_n (\as L)^n$ in the limit $\rho \ll y_c \sim \mc{O}(1)$, which is a typical
feature of mMDT~\cite{Dasgupta:2013ihk}.

\subsection{The midrange region: $\rho \sim y_c \ll 1$}
 
In this region, collinear modes cannot be emitted inside the jet since the jet mass is too small, and they cannot satisfy the mMDT criterion either.
Therefore the collinear contribution genuinely becomes the normalization factor $C_k = \mc{J}_k^{-1}$ like the contribution to 
the plain jet mass distribution in the limit $\rho \ll 1$, as  shown in Eq.~\eqref{facpljm}.
The NLO results for $\mc{J}_{k} (Q^2)$ are given in Eqs.~\eqref{qintj} and \eqref{gintj}. 

Then, as explained previously, the nonzero groomed jet mass can be described by the ultracollinear mode and the csoft mode. Since 
the ultracollinear mode cannot recognize the (ungroomed) jet boundary and the mMDT criterion, the contribution yields the standard jet functions,  
which are given to NLO as 
\begin{align}
J_q (M^2,\mu) = \delta (M^2) +&\frac{\as C_F}{2\pi} \Bigl\{ \delta (M^2) \Bigl[  \frac{3}{2} \ln  \frac{\mu^2}{\Lambda^2} 
+ \ln^2  \frac{\mu^2}{\Lambda^2}+\frac{7}{2} -\frac{\pi^2}{2}\Bigr]  
 -\Bigl[ \Bigl(   2 \ln  \frac{\mu^2}{M^2} +\frac{3}{2} \Bigr) \frac{1}{M^2}\Bigr]_{\Lambda^2} \Bigr\}, \nonumber \\
\label{jetnlo}
J_g (M^2,\mu) = \delta (M^2) + &\frac{\alpha_s}{2\pi} \Bigl\{\delta (M^2) \Bigl[C_A \Bigl( \ln \frac{\mu^2}{\Lambda^2} 
+\frac{67}{18} -\frac{\pi^2}{2}
\Bigr) -\frac{10}{9} T_R n_f +\frac{\beta_0}{2} \ln \frac{\mu^2}{\Lambda^2}\Bigr] \nonumber \\
&- \Bigl[ \Bigl(\frac{\beta_0}{2} + 2C_A \ln \frac{\mu^2}{M^2}\Bigr) \frac{1}{M^2}\Bigr]_{\Lambda^2}\Bigr\}.
\end{align}

The decoupled csoft gluons from the collinear fields form the csoft Wilson lines $Y_{n(\n),\cs}$. Similar to Eq.~\eqref{sqf}, the contribution 
to the groomed jet mass (with an energetic quark) is expressed as 
\be
\label{defs1}
S_{k=q}^I (M_{cs}^2=p_J^+\ell_-,\mu) =  \frac{1}{p_J^+ N_c} \mr{Tr}~\langle 0 | Y_{\n,\cs}^{\dg} Y_{n,\cs}
\delta(\ell_-+\Theta_{\cs} i\partial_- ) Y_{n,\cs}^{\dg} Y_{\n,\cs} |0\rangle_. 
\ee
And $S_g^{I}$ is expressed in terms of the Wilson lines in the adjoint representation. 
Here $\Theta_{\cs}$ represents the mMDT criterion in Eq.~\eqref{mdtc}, which the csoft mode should pass.
The phase space for the csoft mode to pass the criterion is illustrated as a shaded region in Fig.~\ref{fig0}-(b). 
In the shaded region, the csoft function $S_k^I$ contributes to the nonzero groomed jet mass, while the contributions from the remaining region and 
the virtual correction yield the part proportional to $\delta(M^2)$. 

The contribution from the shaded region in Fig.~\ref{fig0}-(b)
involves the IR divergence as $M$ goes to zero. This IR divergence is proportional to $\delta(M^2)$ in the 
$\Lambda$-distribution defined in Eq.~\eqref{lamdad}.
Then the IR divergence in the real emission, appearing in the part with $\delta(M^2)$, is cancelled by 
the virtual contribution. And there remains only the UV divergence.
As a result we find that the renormalized csoft functions to NLO are given as
\begin{eqnarray} \label{soft1}
S^I_k (M^2,\mu) &=& \delta (M^2) +\frac{\as C_k}{2\pi} \Bigl\{ -\delta (M^2) \Bigl(  
\frac{1}{2} \ln \frac{\mu^2}{y_c^2 Q^2}  -\frac{\pi^2}{12}\Bigr)  \nonumber \\
&&+\frac{2}{M^2}  \Theta (M^2 -M_c^2) \ln \frac{\mu^2 Q^2}{(M^2)^2}   
+\Bigl[\frac{2}{M^2}  \Theta (M_c^2 -M^2) \ln \frac{\mu^2}{y_c M^2} \Bigr]_{M_c^2} \Bigr\},
\end{eqnarray}
where $M_c^2 = y_c Q^2$, and we set the upper bound to $\Lambda^2 = M_c^2$ in the $\Lambda$-distribution. 
The results with mMDT, along with those with soft drop in Appendix~\ref{sisd} are new.

Finally the groomed jet mass distribution functions $\Phi_k$ in the limit $\rho\sim y_c \ll 1$ are factorized as
\begin{equation}
\label{factI}
\Phi_k (y_c \ll 1, M_J^2\sim y_c Q^2 ) = \mathcal{C}_k (Q^2, \mu) \int_0^{M_J^2} dM^2 J_k (M^2, \mu) S_k^I (M_J^2 -M^2, \mu).
\end{equation}
The factorization structure is the same as the factorized expression for the plain jet mass in the limit $\rho \ll 1$. [See Eq.~\eqref{facpljm}.] 
What is different from the plain jet mass distribution 
is that the csoft function $S_k^I$ in Eq.~\eqref{factI} is affected by the mMDT criterion. 
Note that both groomed and ungroomed jet mass distributions are scale invariant. 
Therefore the renormalization behavior of $S_k^I$ is the same in both cases, too. It will be explained in detail
in Sec.~\ref{rsgjm}.

Expanding all the terms to NLO in the factorized parts, the groomed jet mass distribution functions 
$\Phi_k^{(1)} (y_c ,M^2)$ at fixed order in $\as$ are given as
\begin{eqnarray}
\label{phiq1}
\Phi_q^{(1)} (y_c,M_J^2) &=& \frac{\as C_F}{2\pi} \Bigl\{ \delta (M_J^2) \Bigl( -\ln^2 y_c -\frac{3}{2} \ln y_c -3 
+\frac{\pi^2}{3}\Bigr)  \\
&+&\frac{1}{M_J^2} \Bigl(2 \ln \frac{Q^2}{M_J^2} -\frac{3}{2}\Bigr) \Theta (M_J^2 -M_c^2)
-\Bigl( 2 \ln y_c + \frac{3}{2}\Bigr) \Bigl[ \frac{\Theta (M_c^2 -M_J^2)}{M^2}\Bigr]_{M_c^2} \Bigr\}, \nonumber \\
\label{phig1}
\Phi_g^{(1)} (y_c, M_J^2) &=& \frac{\as}{2\pi} \Bigl\{ \delta (M_J^2) \Bigl[ C_A \Bigl( -\ln^2 y_c -\frac{67}{18} 
+\frac{\pi^2}{3}\Bigr) 
-\frac{\beta_0}{2} \ln y_c +T_R n_f \frac{13}{9}\Bigr]  \\
&+& \frac{2C_A}{M_J^2} \ln \frac{\mu^2 Q^2}{(M_J^2)^2} \Theta (M_J^2 -M_c^2) -\Bigl( \frac{\beta_0}{2} +2 C_A \ln y\Bigr) 
\Bigl[\frac{1}{M_J^2} \Bigr]_{M_c^2} \Theta ( M_c^2 -M_J^2) \Bigr\}. \nnb
\end{eqnarray}
These are consistent with the results in Eq.~(\ref{phi1}) in the limit $y_c \ll \mc{O}(1)$.

\subsection{The peak region: $\rho \ll y_c \ll 1$}

In the peak region, there are distinct
contributions from the collinear modes with $p_c^2 \sim (p_T^J R)^2$,  the csoft modes
with $p_{\mathrm{cs}}^2 \sim y_c^2 (p_T^J R)^2$, the ultracollinear modes with $p_{\mathrm{uc}}^2 \sim M_J^2$ and finally the 
ucsoft modes with $p_{\mathrm{ucs}}^2 \sim  y_c M_J^2$ to express the groomed jet mass distribution $\Phi_k$.  
The collinear mode cannot radiate inside a jet since $\rho \ll 1$, and the csoft mode cannot satisfy 
$\rho \ll y_c$. Therefore none of these modes contributes to nonzero jet mass, hence these modes contribute only to the normalization.
On the other hand, the ultracollinear and the ucsoft modes can contribute to the nonzero groomed jet mass.

The factorized groomed jet mass distribution functions are written as ($k=q$, $g$)
\begin{equation}
\label{factII}
\Phi_k (y_c\ll 1, M_J^2 \ll y_c Q^2 ) = \mathcal{C}_k (Q^2, \mu) \mathcal{S}^{II}_k (y_c^2 Q^2, \mu) \int_0^{M_J^2} dM^2 
J_k (M^2,\mu) U_k (M_J^2 -M^2, \mu).
\end{equation}
The collinear functions $\mathcal{C}_k$ are the inverse of the integrated jet functions $\mathcal{J}^{-1}_k (Q^2, \mu)$ as in 
Eqs.~\eqref{facpljm} and \eqref{factI}. However, 
$\mathcal{S}_k^{II}$ are the csoft contributions in this region, different from $S_k^I$ in Eq.~\eqref{factI}. 
$J_k$  are the standard jet functions for the ultracollinear mode, that also appear in Eqs.~\eqref{facpljm} and \eqref{factI}. 
Finally $U_k$ are the ucsoft contributions to the jet mass distributions.

The detailed calculation of $\mathcal{S}_k^{II}$ is shown in Appendix~\ref{apc}. 
After the zero-bin subtraction, the NLO results are free of IR divergence and the renormalized results are given as
\begin{equation}
\label{s2nlo}\mc{S}^{II}_k (y_c^2Q^2 ,\mu) =  1+\frac{\as C_k}{2\pi} \Bigl(  \frac{1}{2}
\ln^2 \frac{\mu^2}{y_c^2 Q^2} -\frac{\pi^2}{12}\Bigr).
\end{equation}

The ucsoft functions $U_k$ in Eq.~\eqref{factII} can be defined in a similar way to $S_k^I$ in Eq.~\eqref{defs1}, except that
the Wilson lines are replaced by those with the ucsoft gauge fields and the conditional function $\Theta_{\mathrm{cs}}$ is replaced by 
$\Theta_{\mathrm{ucs}}$ due to 
the different scaling of the ucsoft mode. As a result, the quark ucsoft function is defined as 
\be
\label{defu}
U_{q} (M^2=p_J^+\ell_-,\mu) =  \frac{1}{p_J^+ N_c} \mr{Tr}~\langle 0 | Y_{\n,\ucs}^{\dg} Y_{n,\ucs}
\delta(\ell_-+\Theta_{\ucs} i\partial_- ) Y_{n,\ucs}^{\dg} Y_{\n,\ucs} |0\rangle_. 
\ee
Here $\Theta_{\ucs}=\Theta(- i\partial_+ - y_c p_J^+)$, hence the ucsoft contribution to the nonzero mass comes from the shaded region 
in Fig.~\ref{fig0}-(c). The ucsoft mode scales as Eq.~\eqref{pucs} and cannot recognize the jet boundary. 
The gluon ucsoft function $U_g$ can be similarly defined in terms of the Wilson lines in the adjoint representation.  

Similar to computing $S_k^I$, we can calculate the NLO contributions to $U_k$ employing the $\Lambda$-distribution. 
Then the IR divergences are cancelled by the virtual contributions, and the renormalized results to NLO are given as  
\begin{equation}
\label{Unlo}
U_k (M^2,\mu)  = \delta (M^2) \Bigl[ 1-\frac{\as C_k}{\pi} \Bigl( \frac{1}{2} \ln^2 \frac{\mu^2}{y_c\Lambda^2}
 -\frac{\pi^2}{12} \Bigr) \Bigr]  + \frac{\as C_k}{\pi} \Bigl[ \frac{1}{M^2}   \ln \frac{\mu^2}{y_c M^2}  \Bigr]_{\Lambda^2}.
\end{equation}
By expanding the factorized functions in Eq.~\eqref{factII} to order $\alpha_s$, we verify that the NLO results are consistent with 
the results in Eq.~\eqref{phi1} [or Eqs.~\eqref{phiq1} and \eqref{phig1}] after taking the limit $\rho \ll y_c \ll 1$. 
The ucsoft function is also computed using soft drop, and it is given in Appendix~\ref{apeak}.

\section{Resummed Groomed Jet Mass Distributions\label{rsgjm}}
We perform the resummation of the large logarithms arising from small $y_c$ and $\rho$ in the groomed jet mass distributions. 
Though we will fix $y_c = 0.1$ in numerical analysis, here we regard $y_c$ as a very small number in order to investigate the resummed effects of 
the large logarithms in $y_c$. We have two distinctive small jet-mass
regions: the midrange region $(\rho \sim y_c \ll 1)$ and the peak region $(\rho \ll y_c \ll 1)$. These two regions 
have different factorized structures, hence different resummation effects. 

The resummation of large logarithms can be achieved by solving the RG equation of the factorized
parts. We  first factorize the groomed jet mass distributions, and find the appropriate scales at which the logarithms of the factorized functions 
are minimized. Then each factorized function is evolved by the RG equation to a common factorization scale $\mu_f$.
Combining all the RG evolutions of each factorized yields the resummation
of the large logarithms. 

The groomed jet mass distribution functions in the midrange and peak regions involve double logarithms at NLO. 
So the resummed result  at leading logarithmic (LL) accuracy,  $\sum_{n=0} a_k (\as L^2)^n \sim \exp(L f_0 (\as L))$, is
estimated to be larger than the $\mO(1)$ contributions. Here $L$ represents the large logarithms in small 
$\rho$ or $y_c$. Hence, in order to include $\mO(1)$ contributions we resum the large logarithms up to NLL accuracy, 
which is schematically given as $\sum_{n=0} b_k (\as L)^n \sim \exp(f_1 (\as L))\sim \mO(1)$. 

From now on, we use the dimensionless jet mass variable $\rho = M_{(J)}^2/Q^2$ to consider the dimensionless functions with $\rho$ for 
the jet mass distribution and its factorized functions.\footnote{Throughout this paper 
the dimensionless jet mass variable $\rho$ indicates $M_J^2/Q^2$ in most cases. However, it is sometimes used for expressing a partial jet mass 
squared over $Q^2$ (i.e., $M^2/Q^2$) when we consider the dimensionless factorized functions such as $\bar{J}_k (\rho)$ and $\bar{U}_k (\rho)$.} 
The relation between the dimensionless 
functions $\bar{f}(\rho)$ and the dimensionful functions $f(M^2)$ is given by $\bar{f}(\rho) = Q^2 f(M^2)$, where 
$f=\{\Phi_k,\tilde{J}_k,J_k,S_k^I,U_k\}$. With the dimensionless functions, we can express the $\Lambda$-distributions in the dimensionful 
functions in terms of the standard plus distributions and $\delta(\rho)$. The standard plus distribution is defined as
\be
\int^x_0 d\rho [g(\rho)]_+ h(\rho) = \int^x_0 d\rho g(\rho) h(\rho) - h(0)\int^1_0 d\rho g(\rho), 
\ee
where $h(\rho)$ is an arbitrary function that is smooth at $\rho =0$. The details of the conversion are shown in Appendix~\ref{apb}.

\subsection{Midrange region: $\rho \sim y_c \ll 1$}
The factorization theorem in the midrange region is given in Eq.~(\ref{factI}), and  the factorized 
functions satisfy the following RG equations: 
\be
\label{RGEsI}
\frac{d}{d\ln\mu} \mc{C}_k = \gamma_{\mc{C}}^k~ \mc{C}_k, 
~~~\frac{d}{d\ln\mu} \bar{f}_k (\rho) = \int^{\rho}_0 d\rho' \gamma_f^k (\rho') \bar{f}_k (\rho-\rho'), 
\ee
where $k=q,~g$, and $\bar{f}_k=\bar{J}_k,~\bar{S}_k^I$. 
The anomalous dimensions in general can be expressed as  
\bea
\label{anomc} 
\gamma_{\mc{C}}^{k} &=& A_c \Gamma_C^{k} \ln \frac{\mu^2}{Q^2}+\hat{\gamma}_c^k, \\ 
\label{anomj}
\gamma_{J}^{k} (\rho) &=& \delta(M^2) \Bigl(A_j \Gamma_C^{k} \ln \frac{\mu^2}{Q^2}+\hat{\gamma}_j^k\Bigr)-\kappa_j A_j 
\Gamma_C^{k}\Bigl[\frac{1}{\rho}\Bigr]_{+}\ , \\
\label{anoms1}
\gamma_{S^I}^{k} (\rho) &=& \delta(\rho) \Bigl(A_s \Gamma_C^{k} \ln 
\frac{\mu^2}{Q^2}+\hat{\gamma}_{s1}^k\Bigr)
-\kappa_s A_s \Gamma_C^{k}\Bigl[\frac{1}{\rho}\Bigr]_{+}\ ,
\eea
where $\Gamma_{C}^k = \sum_{n=0} \Gamma_{n}^k(\as/4\pi)^{n+1} $ are the cusp anomalous
dimensions~\cite{Korchemsky:1987wg,Korchemskaya:1992je}, and the first two coefficients are given by 
\be
\Gamma_{0}^k = 4C_k,~~~\Gamma_{1}^k = 4C_k \Biggl[\Bigl(\frac{67}{9}-\frac{\pi^2}{3}\Bigr) C_A - \frac{10}{9} n_f\Biggr]\ ,
\ee
with $C_q = C_F$ and $C_g = C_A$.
From the NLO results in Eqs.~\eqref{qintj}, \eqref{gintj}, (\ref{jetnlo}), and (\ref{soft1}), we extract the set of the coefficients 
$\{A_c,A_{j},A_s, \kappa_{j},\kappa_s\} = \{-1,2,-1,1,2\}$. And the noncusp anomalous dimensions at NLL accuracy are 
given by $\hat{\gamma}_{c}^q=-3\as C_F/(2\pi)$, $\hat{\gamma}_{c}^g=-\as \beta_0/(2\pi)$,
$\hat{\gamma}_j^k=-\hat{\gamma}_{c}^k$, and $\hat{\gamma}_{s1}^k=0$.

We find that the factorization structure for the groomed jet mass distribution is the same as that for the plain jet mass distribution 
in the limit $\rho \ll 1$. (See Ref.~\cite{Idilbi:2016hoa}.)
In this limit the grooming effects (represented by $y_c$) appear only in the csoft function. 
However, the RG behavior for the csoft function should be the same as the plain jet mass since both jet mass distributions 
are scale invariant and the collinear function $\mc{C}_k$ and the jet function $J_k$ are the same in both cases.

The RG equations can be solved by following the conventional methods using the Laplace 
transform~\cite{Neubert:2005nt,Becher:2006nr}, and the resummed result at NLL accuracy is written as 
\bea 
\label{rPhik1}
\bar{\Phi}_k (\rho\sim y_c) =  \Phi_k (M_J^2) \cdot Q^2 &=& \exp[\mc{M}^I_k 
(\mu_{\cc},\mu_{\uc},\mu_{\cs})]~\mc{C}_k(Q,\mu_{\cc})  \\
&&\times\hat{J}_k \Bigl[\ln\frac{\mu_{\mr{uc}}^2}{Q^2}-\partial_{\eta_1}\Bigr] 
\hat{S}^I_k \Bigl[\ln\frac{\mu_{\mr{cs}}^2}{Q^2}-2\partial_{\eta_1}\Bigr]
\frac{e^{-\gamma_E\eta_1}}{\Gamma(\eta_1)} \rho^{-1+\eta_1},\nnb
\eea
where $\bar{\Phi}_k(\rho)$ is the dimensionless jet mass distribution with $\rho = M_J^2/Q^2$, and $Q= p_T^J R$. And $\hat{J}_k$ and $\hat{S}_k$ 
in Eq.~(\ref{rPhik1}) are the Laplace transforms of the dimensionless functions $\bar{J}_k(\rho)$ and $\bar{S}_k^I (\rho)$, 
which are given to NLO as 
\bea
\hat{J}_q [L] &=& 1+\frac{\as C_F}{2\pi} \Bigl(\frac{7}{2}-\frac{\pi^2}{3}+\frac{3}{2}L+L^2\Bigr),\\
\hat{J}_g [L] &=& 1+\frac{\as C_A}{2\pi} \Bigl(\frac{67}{18}-\frac{5n_f}{9 C_A}-\frac{\pi^2}{3}
+\frac{\beta_0}{2C_A}L+L^2\Bigr),\\
\hat{S}^I_k [L] &=& 1+\frac{\as C_k}{2\pi} \Bigl(-\frac{1}{2}L^2-\frac{\pi^2}{4}\Bigr).
\eea
In Eq.~(\ref{rPhik1}),  $\eta_1 = 2a[\Gamma_C^k](\mu_{\uc},\mu_{\cs})$ and it is positive for $\mu_{\uc} > \mu_{\cs}$. 
Since $\Phi_k$ is scale invariant, the factorization scale $\mu_f$ does not appear in Eq.~(\ref{rPhik1}).  

The exponent $\mc{M}^I_k$ at NLL accuracy is written as 
\bea
\label{expf1}
\mc{M}^I_k (\mu_{\cc},\mu_{\uc},\mu_{\cs}) &=& -2 S_{\Gamma}^k (\mu_{\uc},\mu_{\cc})
-2 S_{\Gamma}^k (\mu_{\uc},\mu_{\cs}) \\
&&+ \ln\frac{\mu_{\uc}^2}{Q^2}\Bigl(a[\Gamma_C^k](\mu_{\cc},\mu_{\uc})
+a[\Gamma_C^k](\mu_{\cs},\mu_{\uc})\Bigr)
+a[\hat{\gamma}_j^k](\mu_{\cc},\mu_{\uc}). \nnb
\eea
Here $S_{\Gamma}^k$ and $a[f]$ are defined as 
\be
S_{\Gamma}^k (\mu_1,\mu_2) = \int^{\alpha_1}_{\alpha_2} \frac{d\as}{b(\as)} \Gamma_{C}^k(\as) \int^{\as}_{\alpha_1}
\frac{d\as'}{b(\as')},~~~a[f](\mu_1,\mu_2) = \int^{\alpha_1}_{\alpha_2} \frac{d\as}{b(\as)} f(\as),
\ee
where $\alpha_{1,2} \equiv \as (\mu_{1,2})$ and $b(\as)=d\as/d\ln\mu$ is the QCD beta function.

\subsection{Peak region: $\rho \ll y_c \ll 1$}
The groomed jet mass distribution is factorized in terms of the 
collinear~($\mc{C}_k$), csoft~($\mc{S}_k^{II}$), 
ultracollinear~($J_k$), and ucsoft~($U_k$) functions, as shown in Eq.~(\ref{factII}).
Here the collinear functions $\mc{C}_k$ and the jet functions $J_k$ for the ultracollinear modes are the same as those in 
the midrange region. The RG equations for $\mc{S}_k^{II}$ and the dimensionless function $\bar{U}_k$ are given by 
\be
\label{RGEsII}
\frac{d}{d\ln\mu} \mc{S}^{II}_k = \gamma_{\mc{S}^{II}}^k \mc{S}^{II}_k, 
~~~\frac{d}{d\ln\mu} \bar{U}_k (\rho) = \int^{\rho}_0 d\rho' \gamma_U^k (\rho') \bar{U}_k (\rho-\rho'). 
\ee
And the anomalous dimensions are written as 
\bea
\label{anoms2} 
\gamma_{\mc{S}^{II}}^{k} &=& A_{s2} \Gamma_C^{k} \ln \frac{\mu^2}{y_c^2 Q^2}+\hat{\gamma}_{s2}^k, \\ 
\label{anomu}
\gamma_{U}^{k} (\rho) &=& \delta(\rho) \Bigl(A_u \Gamma_C^{k} \ln \frac{\mu^2}{y_cQ^2}+\hat{\gamma}_u^k\Bigr)
-\kappa_u A_u \Gamma_C^{k}\Bigl[\frac{1}{\rho}\Bigr]_{+}\ . 
\eea
The coefficients $\{A_{s2},A_u,\kappa_u\}=\{1,-2,1\}$ are obtained from the NLO results in 
Eqs.~(\ref{s2nlo}) and (\ref{Unlo}), and the noncusp anomalous dimensions 
at NLL accuracy are  $\hat{\gamma}_{s2}^k=\hat{\gamma}_u^k=0$. The scale invariance of $\Phi_k(\rho\ll y_c)$ is 
guaranteed by the following relations: 
\be
A_{c}+A_{s2}+A_j+A_u=0,~~~\kappa_j A_j + \kappa_u A_u = 0,~~~\hat{\gamma}_c^k+\hat{\gamma}_j^k=0.
\ee

By evolving the factorized functions from their own scales to the factorization
scale $\mu_f$ from Eq.~(\ref{RGEsII}), the resummed results are written as 
\bea 
\label{rPhik2}
\bar{\Phi}_k (\rho\ll y_c)   &=& \exp[\mc{M}^{II}_k(\mu_{\cc},\mu_{\cs},\mu_{\uc},\mu_{\ucs})]
~\mc{C}_k(Q,\mu_{\cc})\mc{S}_k^{II}(y_cQ,\mu_{\cs})  \\
&&\times\hat{J}_k \Bigl[\ln\frac{\mu_{\mr{uc}}^2}{Q^2}-\partial_{\eta_2}\Bigr] \hat{U}_k
\Bigl[\ln\frac{\mu_{\mr{\ucs}}^2}{Q^2}-\partial_{\eta_2}\Bigr]
\frac{e^{-\gamma_E\eta_2}}{\Gamma(\eta_2)} \rho^{-1+\eta_2},\nnb
\eea
where $\eta_2 = 2a[\Gamma_C^k](\mu_{\uc},\mu_{\ucs})$. And $\hat{U}_k$ are the Laplace transforms of the dimensionless ucsoft 
functions $\bar{U}_k(\rho)$, which are given to NLO as 
\be
\hat{U}_k [L] = 1-\frac{\as C_k}{2\pi} L^2. 
\ee
The exponent $\mc{M}_{k}^{II}$ is written as 
\bea
\label{expf2}
\mc{M}^{II}_k (\mu_{\cc},\mu_{\cs},\mu_{\uc},\mu_{\ucs}) &=& 2 S_{\Gamma}^k (\mu_{\cc},\mu_{\cs})-4 S_{\Gamma}^k
(\mu_{\uc},\mu_{\ucs}) + \ln\frac{\mu_{\cc}^2}{Q^2}a[\Gamma_C^k](\mu_{\cc},\mu_{\cs})\\
&-&2\ln\frac{\mu_{\uc}^2}{Q^2}a[\Gamma_C^k](\mu_{\uc},\mu_{\ucs})
+2 \ln y_c a[\Gamma_C^k](\mu_{\cs},\mu_{\ucs}) +a[\hat{\gamma}_j^k](\mu_{\cc},\mu_{\uc}). \nnb
\eea

\subsection{Nonglobal Logarithms}
NGLs~\cite{Dasgupta:2001sh,Banfi:2002hw} arise when gluons are radiated across the jet boundary, and contribute to 
a jet observable with the phase space constrained by the boundary. 
Although the leading NGLs begin to appear at two loop, the perturbative series is schematically given as 
$\sum_{n=2} b_{\rm{NG}}^n (\as L_{\mr{NGL}})^n$, and hence contributes at NLL accuracy.
Especially when there is a large energy difference between the gluons across the boundary, large NGLs appear.
In the effective theory approach, when there are multiple modes to resolve the jet boundary and there is a large energy
hierarchy among them, there appear large NGLs. 

The groomed jet mass in the midrange region ($\rho \sim y_c \ll 1$) is a NGL observable since  
the collinear and csoft modes can resolve the jet boundary and the csoft mode can contribute to the groomed jet mass. 
In the peak region ($\rho \ll y_c \ll 1$), although the NGLs can be generated by the collinear and csoft modes, they do not 
affect the jet mass directly since the collinear and csoft modes contribute only to the normalization of the jet mass distribution.

For the plain jet mass distributions, the contribution of the NGLs is sizable around the peak region,
but decreases rapidly away from the peak. However, we expect that the groomed jet mass is not affected substantially by 
the NGLs. The groomed jet mass in the peak region is not a NGL observable at all, and the NGL contribution in the 
midrange region is quite suppressed since this region is far away from the peak.   

It is remarkable that the resummed result of the leading NGLs for a narrow isolated jet would take the same form as that for 
the hemisphere jet mass since the generating mechanism of the NGLs is similar~\cite{Banfi:2010pa,Dasgupta:2012hg}.
Therefore in order to estimate the NGL contribution to the groomed jet mass, we might be able to use the resummed formula 
of the leading NGLs in the large $N_c$ limit for the hemisphere jet mass~\cite{Dasgupta:2001sh}, which is written as 
\be\label{RNGL}
\Delta^{k=q,g}_{\mr{NG}} (\mu_{1},\mu_{2}) = \exp\Biggl(-C_A C_k \frac{\pi^2}{3} \Bigl(\frac{1+(at)^2}{1+(bt)^c}\Bigr) 
t^2 \Biggr)\ ,
\ee
where 
\be
t=\frac{1}{\beta_0} \ln{\frac{\as(\mu_{2})}{\as(\mu_{1})}} \sim -\frac{1}{\beta_0}\ln \Bigl(1-\frac{\beta_0}{4\pi} \as(\mu_1) 
\ln \frac{\mu_1^2}{\mu_{2}^2}\Bigr)\ .
\ee 
Here the fit parameters from the Monte Carlo simulation are given by $a=0.85 C_A,~b=0.86C_A$, and
$c=1.33$~\cite{Dasgupta:2001sh}. In our analysis we set $(\mu_1,\mu_2)=(\mu_{\cc},\mu_{\cs}^I)$ for the midrange 
region and $(\mu_1,\mu_2)=(\mu_{\cc},\mu_{\cs}^{II})$ for the peak region. For the NLL-resummed results  
(with the fixed NLO), the contributions of the NGLs in Eq.~(\ref{RNGL}) is multiplicative in Eq.~(\ref{rPhik1}) or
Eq.~(\ref{rPhik2}).

\section{Numerical Results\label{num}}
We present the numerical analysis on the groomed jet mass distributions $\bar{\Phi}_k (\rho)$. In fact, they are not physical observables, and
for phenomenology, the ratio of the scattering cross sections in Eqs.~(\ref{defjff}) and (\ref{jsec}) with respect to the jet mass  
should be considered by summing over all the contributions combined with the parton distribution functions. 
For example, the phenomenological analysis in the peak region has been performed in Ref.~\cite{Kang:2018jwa}. However, our main focus
is the theoretical issue on how to implement the grooming method to the jet mass distribution, and how the resummation on $\rho$ and $y_c$
affects the jet mass distribution over a wide range of $\rho$.
Therefore the purpose of 
the numerical analysis is to offer theoretical understanding on how the groomed jet mass distributions for the quark- and 
gluon-initiated jets behave with grooming and how the results in different regions are affected by the resummation. 
A complete phenomenological analysis is beyond the scope of this paper, and will be performed in future work.

\begin{figure*}[t]
\begin{center}
\includegraphics[width=15cm]{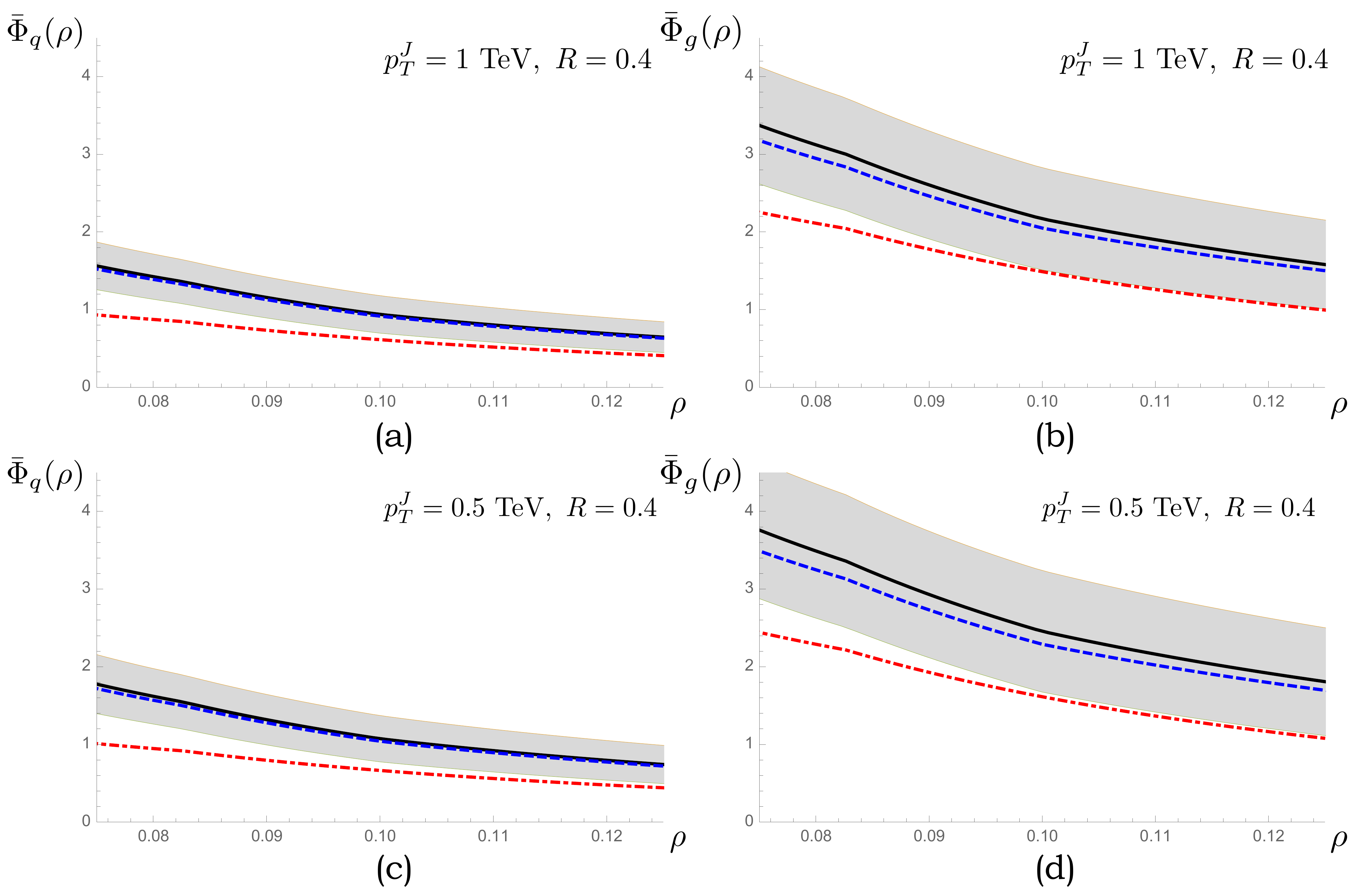}
\end{center}
\vspace{-0.6cm}
\caption{\label{fig1} \baselineskip 3.0ex 
Groomed jet mass distributions multiplied by $\rho$ for a quark-initiated jet [Fig.~\ref{fig1}-(a), (c)] and a gluon-initiated 
jet [Fig.~\ref{fig1}-(b), (d)] with $(p_T^J,R)=(1~\mr{TeV},0.4)$ and $(p_T^J,R)=(0.5~\mr{TeV},0.4)$ in the midrange region
($\rho\sim y_c \ll 1$). Here $\rho = M_J^2/(p_T^JR)^2$. The black thick lines denote the NLL resummed results with the fixed 
NLO corrections ($\mr{NLL_{G}+NLO}$). The blue dashed lines denote the resummed results including the NGLs
($\mr{NLL_{G+NG}+NLO}$). The red dot-dashed lines are the results at NLO without resummation. Gray bands at
$\mr{NLL_{G}+NLO}$ show the uncertainties under the scale variations from $\mu_{c,uc,cs} = 2\mu_{\cc,\uc,\cs}^{0}$ 
to $\mu_{\cc,\uc,\cs}^{0}/2$. 
} 
\end{figure*}

For numerical analysis, we set $y_c = 0.1$. First of all, let us consider 
the behavior of the groomed jet mass distribution 
in the new region, which we call the midrange region ($\rho \sim y_c \ll 1$)
in Fig.~\ref{fig1}. Here black thick lines show the results of resumming large
logarithms of small $\rho$ and $y_c$ at NLL accuracy with  the fixed NLO corrections ($\mr{NLL_G+NLO}$), where
$\mr{NLL_G}$ means the resummation without NGLs. These results can be directly obtained from Eq.~(\ref{rPhik1}). 
We set the default collinear, ultracollinear, and csoft scales as 
$(\mu^0_{\cc},\mu^0_{\uc},\mu^{I,0}_{\cs})=(Q, \sqrt{\rho} Q, \sqrt{\rho y_c} Q)$. Though 
the resummed result is independent of $\mu_f$, there is arbitrariness in setting the default scale for each factorized
function. The uncertainties when each scale varies from $2\mu_i^0$ to $\mu_i^0/2$ separately are shown as gray bands 
in Fig.~\ref{fig1}. 

Compared to the fixed NLO results without resummation (red dot-dashed lines in Fig.~\ref{fig1}), the resummed results
are significantly enhanced by $50-80~\%$. Since the jet mass distribution for 
a gluon-initiated jet is broad and decreases slowly away from the peak, the distribution for the 
gluon-initiated jet is dominant over the quark-initiated jet in this region. Note that the contributions of the NGLs 
(blue dashed lines in in Fig.~\ref{fig1}) are small even though the jet mass in this region is a nonglobal variable.  

\begin{figure*}[t]
\begin{center}
\includegraphics[width=15cm]{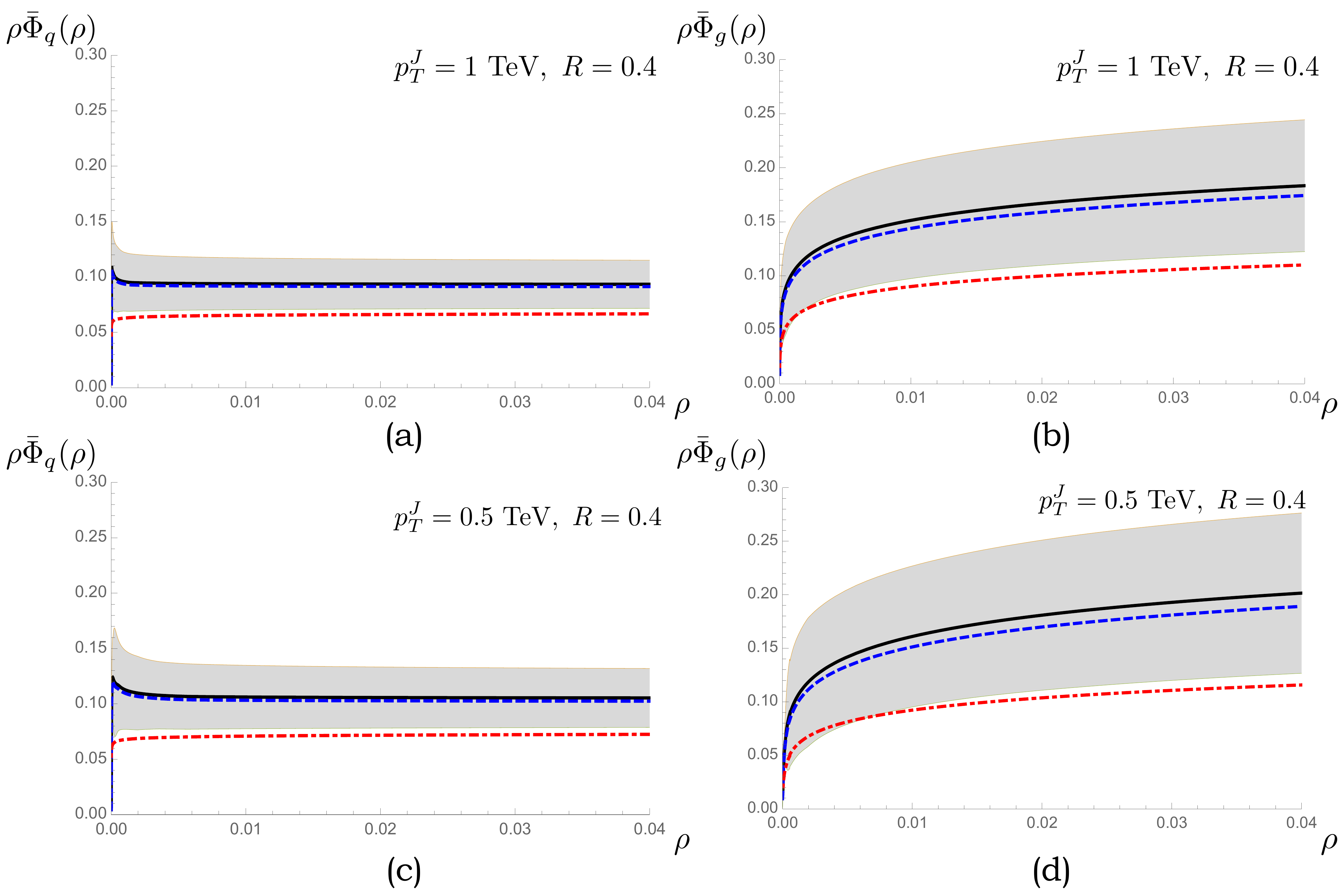}
\end{center}
\vspace{-0.6cm}
\caption{\label{fig2} \baselineskip 3.0ex 
Groomed jet mass distributions multiplied by $\rho$ in the peak region ($\rho\ll y_c$). Here Fig.~\ref{fig2}- (a,c) are for the quark-initiated jet and Fig.~\ref{fig2}-(b,d) for the gluon-initiated jet. 
The black thick lines denote the NLL resummed results with the fixed NLO corrections ($\mr{NLL_{G}+NLO}$) and the 
blue dashed lines denote the resummed results including NGLs ($\mr{NLL_{G+NG}+NLO}$). The red dot-dashed lines are 
the results resumming $\ln \rho$ at LL accuracy. Gray bands for the results at $\mr{NLL_{G}+NLO}$ are the 
uncertainties under the scale variations from 
$\mu_{\cc,\cs,\uc,\ucs} = 2\mu_{\cc,\cs,\uc,\ucs}^{0}$ to $\mu_{\cc,\cs,\uc,\ucs}^{0}/2$. 
} 
\end{figure*}

In Fig.~\ref{fig2}, we show the groomed jet mass distributions (multiplied by $\rho$) in the peak region ($\rho \ll y_c\ll 1$).
Here black thick lines are our default results with the accuracy of $\mr{NLL_G+NLO}$ using Eq.~(\ref{rPhik2}). 
The default scales for the factorized functions are chosen as 
$(\mu^0_{\cc},\mu_{\cs}^{II,0},\mu^0_{\uc},\mu^{0}_{\ucs})=(Q, y_c Q, \sqrt{\rho} Q, \sqrt{\rho y_c} Q)$.
We estimate the uncertainties by varying the scales from $2\mu_i^0$ to $\mu_i^0/2$ separately, and they are shown 
as gray bands in Fig.~\ref{fig2}. To avoid the Landau pole as $\rho$ goes to zero, we introduce a small, fixed
point, $\rho_0$. Then, in the region $\rho <\rho_0$, we make the ucsoft scale freeze 
at a value slightly above the Landau pole. This is implemented by using the scale profile as 
\be
\label{ucspf} 
\mu_{\ucs}^{\mr{pf}} = 
\left\{
\begin{array}{rl} 
& Q\sqrt{y_c\rho}~~~\mr{if}~\rho \ge \rho_0,  \\ 
& \mu_{\mr{min}} + a  Q \sqrt{y_c \rho^3} ~~~\mr{if}~\rho < \rho_0,   \end{array} \right.
\ee
where we set $\mu_{\mr{min}} = 0.5~\mr{GeV}$. And $\rho_0$ and $a$ are determined for $\mu_{\ucs}^{\mr{pf}}$ to be
smooth and continuous at $\rho_0$. Accordingly, the ultracollinear scale profile is given as 
$\mu_{\uc}^{\mr{pf}} = \mu_{\ucs}^{\mr{pf}}/\sqrt{y_c}$.  

The NGL contributions to the resummed results (blue dashed lines in Fig.~\ref{fig2}) come from the scale deviation between 
$\mu_{\cc}$ and $\mu_{\cs}^{II}$, hence the NGLs in the peak region takes the form of $\ln y_c$ and it affects the 
normalization of the distributions. As in the case of the midrange region, the effects are tiny, and especially negligible for 
a quark-initiated jet.  

Red dot-dashed lines in Fig.~\ref{fig2} are the results of resumming $\ln \rho$ only with the accuracy of
LL+NLO,\footnote{
Since the results at NLO in Eq.~(\ref{srho}) involve only single logarithms, the resummed results at LL accuracy can be
estimated as $\mO(1)$ and comparable to the results at NLL accuracy shown in Eqs.~(\ref{rPhik1}) and (\ref{rPhik2}).} 
which are based on the factorized results in Eq.~(\ref{srho}) reflecting the limit $\rho \ll y_c \sim \mc{O}(1)$. 
Comparing them to the default results (black thick lines in Fig.~\ref{fig2}), we see large deviations, which indicate that 
the resummation of $\ln y_c$ give rise to a significant enhancement in the peak region.  The enhancement from resumming on 
$\ln y_c$ is persistent from the midrange region to the peak region.

\begin{figure*}[t]
\begin{center}
\includegraphics[width=16cm]{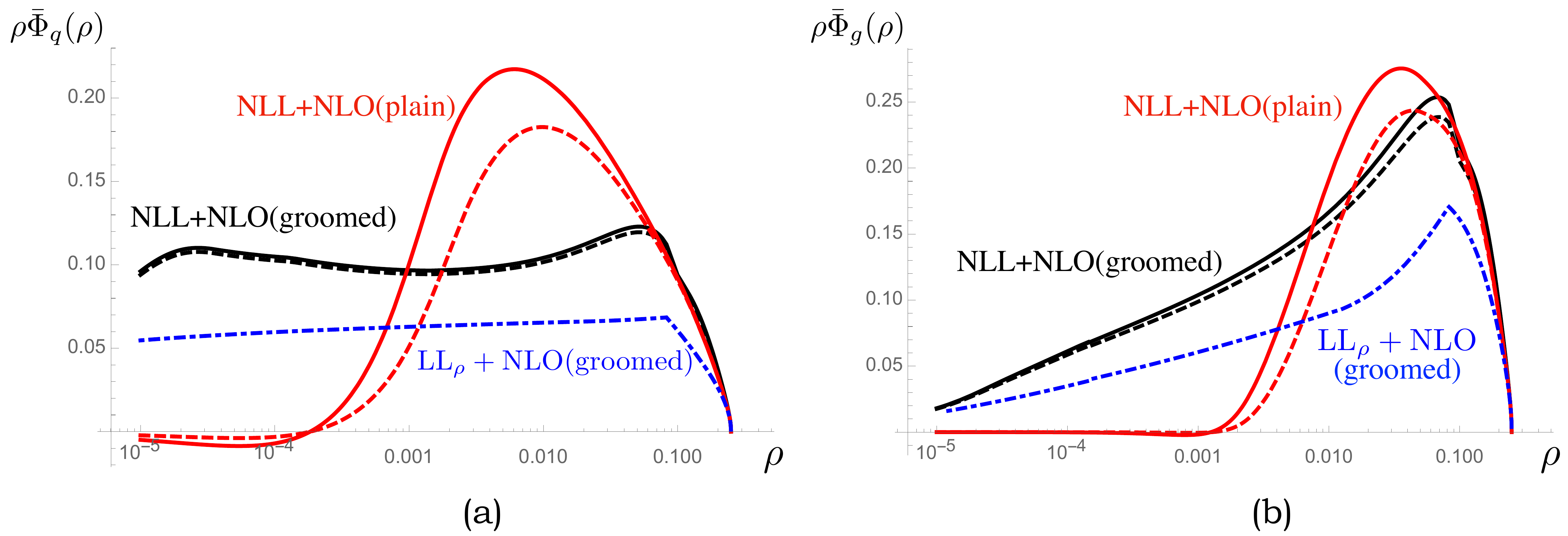}
\end{center}
\vspace{-0.6cm}
\caption{\label{fig3} \baselineskip 3.0ex 
Groomed jet mass distributions multiplied by $\rho$ in the full range of $\rho$ for a quark-initiated jet [Fig.~\ref{fig3}-(a)] 
and a gluon-initiated jet [Fig.~\ref{fig3}-(b)] with $p_T^J=1~\mr{TeV}$ and $R=0.4$. Here black solid (dashed) lines are 
the groomed jet mass distributions at the accuracy of $\mr{NLL_G+NLO~(NLL_{G+NG}+NLO)}$. Red solid (dashed) lines 
are the plain jet mass distributions at the accuracy of $\mr{NLL_G+NLO~(NLL_{G+NG}+NLO)}$. Blue dot-dashed lines denote
the groomed jet mass distributions with resummation of only $\ln \rho$ at LL accuracy. 
} 
\end{figure*}

In Fig.~\ref{fig3} we finally show the groomed jet mass distributions (multiplied by $\rho$) in the full range of $\rho$. 
Here black solid (dashed) lines are fully resummed results for small $\rho$ and $y_c$ at NLL accuracy without (with) 
NGL effects, and they are illustrated by combining the tail ($\rho \gg y_c$), midrange ($\rho \sim y_c$) and the 
peak regions~($\rho\ll y_c$). Here the distributions for the tail region are the same as the plain jet mass 
distributions without resummation. 

In combining the groomed jet mass distributions in three different regions, we interpolate around $\rho \sim 0.08$ 
($\rho \sim 0.12$) for smooth connection between the distributions in the midrange and the peak (tail) 
regions.\footnote{\baselineskip 3.0 ex The interpolation points are somewhat arbitrary. 
For example, if we try interpolation around $\rho \sim 0.06$ instead of $\rho \sim 0.08$, the shapes of the combined
distributions vary slightly giving some uncertainties, which need to be fixed by experimental measurements.} 
Compared with the groomed distributions only with the resummation on $\ln \rho$ (blue dot-dashed lines 
in Fig.~\ref{fig3}), we observe that the resummation of the large logarithm $\ln y_c$ yields a significant enhancement through all the
regions.

The red solid (dashed) lines in Fig.~\ref{fig3} are the plain jet mass distributions at the accuracy of NLL+NLO without (with) 
the NGL effects. Compared with the fully resummed groomed distributions, we see that the NGL effects are 
quite suppressed by the grooming process. Also, especially for the quark-initiated jet, we see that the jet mass distribution 
for $\rho \sim [10^{-3},0.1]$ is affected by the grooming. However, for the gluon-initiated jet, the grooming is not so
effective. The gluon jet distributions are usually broad and have a relatively thick tail region. Hence the grooming parameter
$y_c =0.1$ might not be large enough to suppress multiple (collinear-)soft gluon radiations.  

\section{Conclusion\label{conc}}
We have investigated the factorization of the groomed jet mass distribution in a wide range of the jet mass in the 
effective-theory approach. There are distinct modes in the tail, midrange, and peak regions contributing to each factorized
part. In the tail region, the collinear modes are enough to describe the groomed jet function, which coincides with the ungroomed jet function. 
In the midrange region with 
small jet mass, we need the ultracollinear and csoft modes  additionally. In the peak region with very small jet mass, 
the ucsoft modes are also required. We apply the effective theories appropriate in these regions to obtain
the factorized groomed jet mass distributions and resum the large logarithms on $\rho$ and/or $y_c$. By combining all 
the results, we are able to have a bird's-eye view of the groomed jet mass distribution over the whole range of the jet mass.

The main issue is to implement the grooming procedure in theoretical calculations systematically. We have focused on how to treat the remaining
particles, especially a single particle in a jet when those particles which fail the grooming criterion are removed from the jet. The grooming, rather than
the tagging, is chosen such that
the remaining particles contribute to the groomed jet mass even when the grooming criterion is not satisfied. This prescribes the theoretical 
computation at NLO. When the criterion fails, the remaining single particle contributes to the $\delta (M_J^2)$ part, which includes the IR divergence
in real emissions. However, it is cancelled by the virtual corrections, which also contribute to the $\delta (M_J^2)$ part.
Therefore the groomed jet mass distribution is IR safe and starts from $\alpha_s^0$. 
We focus on the theoretical issues here, and a detailed phenomenological analysis following the experimental setup closely, along with the inclusion of 
the nonperturbative effects will be considered in future work.

In Ref.~\cite{Frye:2016aiz}, the factorization for groomed jet substructure was considered at NNLL accuracy.  They focus on the factorization of the 
hemisphere jet mass distributions, in which the detailed ingredients of the factorizations are different. But the line of reasoning leading to the factorization
is similar to the factorization in the peak region in our paper. In Refs.~\cite{Marzani:2017mva,Marzani:2017kqd}, 
a phenomenological analysis is performed on the groomed jet mass distributions in mMDT and soft drop in QCD with the estimation of nonperturbative effects. 
The approach of Ref.~\cite{Kang:2018jwa} is closely related to our analysis of the peak region.  

The important features of our paper compared to previous literature are the following: First, we elaborate on how to implement grooming theoretically in the 
effective-theory approach. Different modes are identified with different momentum scaling, and the overlap regions between different modes are 
disentangled using the 
zero-bin subtraction. The UV and IR divergences are carefully treated to see if they are separated. After nontrivial computation and consistency check,
we find that the grooming procedure yields IR-safe factorized parts, while the tagging procedure does not at NLO. 
Using grooming, all the factorized parts are IR finite, and
we can apply the RG equations to each of them to resum large logarithms. We here emphasize again that the IR safety or, at least no mixture of IR
and UV divergences (as in the case of the parton distribution functions) is the essential requirement to construct and solve the RG equations.

Second, in order to scan all the possible range of the jet mass, we include the midrange region with $\rho \sim y_c \ll 1$. It is located 
between the peak and the tail regions. In this region, the ultracollinear, csoft and collinear modes give factorized contributions, which are different from
those in other regions. The new results of the factorized groomed jet mass distribution using mMDT and soft drop are presented, and it turns out that
the resummation on $\ln y_c$, as well as $\ln \rho$, gives appreciable enhancements which persists in the peak region. 
In the numerical analysis, the effect of the resummation on $\ln \rho$ and $\ln y_c$ is appreciable 
though we put $y_c =0.1$. This enhances the  groomed jet mass distribution by about 50 -- 80\% compared to the result 
without resummation on $\ln y_c$. And nonglobal logarithms are negligible, which is the characteristic of mMDT and soft drop.   
 
The study of jet substructure has become a mature subject along with the concurrent experimental analysis. The jet substructure can be investigated in various observables
other than the jet mass distribution. They can be probed using higher-order contributions, which include nonglobal and clustering logarithms, or considering
the behavior of the signal jets. The 
analytical comparison of the jet structure between the QCD jets and the signal jets will be a cornerstone to discover new physics through the study of jets. These features
will be explored in future work, based on the approaches presented in this paper.

\appendix

\section{List of all the functions in the groomed mass distribution functions\label{apa}}
Here we list all the functions, which appear in the text. Throughout this section, $y$ denotes the grooming parameter $y_c$. Also all the results in this section can be applied to the case of the soft drop with $\beta=0$ when $y_c$ is replaced with $\zc/(1-\zc)$. 

The functions $I_q (y)$ and $I_g (y)$ in Eq.~(\ref{phi1}) are given by
\begin{eqnarray} \label{iqg}
I_q (y) &=&  \ln^2 y +\frac{3}{2} \ln y +\frac{3 (1-y)}{1+y} -2 \ln y \ln (1+y) + 2 \mathrm{Li}_2 
\Bigl(\frac{y}{1+y}\Bigr) -2 \mathrm{Li}_2 \Bigl(\frac{1}{1+y}\Bigr),  \nonumber \\
I_g (y) &=& \ln^2 y +\frac{11}{6} \ln y + \frac{1-y}{18 (1+y)^3}(67+130y +67y^2) -2 \ln y \ln (1+y)   \\
&&+ 2 \mathrm{Li}_2 \Bigl(\frac{y}{1+y}\Bigr) -2 \mathrm{Li}_2 \Bigl(\frac{1}{1+y}\Bigr) 
+\frac{T_R n_f}{C_A} \Bigl[ -\frac{1}{3}\ln y -\frac{1-y}{18(1+y)^3} (13+22y +13y^2)\Bigr]. \nonumber
\end{eqnarray}
In the limit of small $y$, they become
\begin{eqnarray} \label{izero}
I_q (y)  &\rightarrow& \ln^2 y +\frac{3}{2} \ln y +3-\frac{\pi^2}{3},  \nonumber\\
I_g (y) &\rightarrow& \ln^2 y +\frac{11}{6}\ln y +\frac{67}{18} -\frac{\pi^2}{3}+ \frac{T_R n_f}{C_A} 
\Bigl( -\frac{1}{3}\ln y -\frac{13}{18}\Bigr).
\end{eqnarray}

The functions $f_q(w)$ and $f_g(w)$ are given as
\begin{eqnarray} \label{fqg}
f_q (w) &=& -\frac{3}{2}w +2\ln \frac{1+w}{1-w}, \nonumber\\
f_g (w) &=& -\frac{w}{12} (21+w^2) + 2\ln \frac{1+w}{1-w} +\frac{T_R n_f}{C_A} \frac{w}{12}(3+w^2).
\end{eqnarray}
The functions $g_q (y)$ and $g_g (y)$ are given as
\begin{eqnarray} \label{gqg}
g_q (y)&=& \frac{3}{2} -\frac{3}{1+y}-2 \ln y, \nonumber \\
g_g (y) &=& -\frac{1-y}{6(1+y)^3} (11+ 20y+11y^2) -2 \ln y\Bigr) +\frac{T_R n_f}{C_A} \frac{1-y^3}{3(1+y)^3}.
\end{eqnarray}
In the limit $y\rightarrow 0$, they approach
\begin{equation} \label{gzero}
g_q (y)  \rightarrow  -2 \ln y -\frac{3}{2}, \ \
g_g (y)   \rightarrow    -2 \ln y -\frac{11}{6} + \frac{1}{3}\frac{T_R n_f}{C_A}. 
\end{equation}

The functions $h_q (y)$ and $h_g (y)$ shown in Eqs.~(\ref{tagjm}) and (\ref{hard2}) are given as
\begin{eqnarray}\label{hqg}
h_q (y) &=& -\frac{7}{2} \frac{1-y}{1+y}  +\ln^2 y - 2\ln y \ln (1+y) -\frac{3y}{1+y} \ln y -\frac{3(1-y)}{1+y} 
\ln (1+y) \nonumber \\
&&+2 \mathrm{Li}_2 \Bigl(\frac{1}{1+y}\Bigr) - 2 \mathrm{Li}_2 \Bigl(\frac{y}{1+y}\Bigr),  \nonumber \\
h_g (y) &=& -\frac{1-y}{18(1+y)^3} (67+30y +67y^2) +\ln^2 y-2 \ln y \ln (1+y) \nonumber \\
&&-\frac{y}{3 (1+y)^3} (12+21 y + 11y^2) \ln y -\frac{1-y}{3(1+y)^3} (11 + 20y + 11y^2) \ln (1+y) \nonumber \\
&&+2 \mathrm{Li}_2 \Bigl(\frac{1}{1+y}\Bigr) - 2 \mathrm{Li}_2 \Bigl(\frac{y}{1+y}\Bigr)  \nonumber \\
&&+ \frac{T_R n_f}{C_A} \Bigl[ \frac{y}{3(1+y)^3} (3 + 3 y + 2y^2) \ln y +\frac{1-y^3}{9(1+y)^3} \Bigl(5 
+ 6\ln (1+y)\Bigr) \Bigr],
\end{eqnarray}
and their limiting forms for small $y$ are given as
\begin{equation}
h_q (y) \rightarrow -\frac{7}{2} +\frac{\pi^2}{3} +\ln^2 y,   \ \
h_g (y) \rightarrow -\frac{67}{18} +\frac{\pi^2}{3} +\ln^2 y +\frac{5}{9}\frac{T_R n_f}{C_A}.
\end{equation}

\section{Conversion of the $\Lambda$-distribution to the standard plus distribution\label{apb}}
The jet mass distributions and its factorized functions such as the standard jet function and the csoft and ucsoft functions 
with the $\Lambda$-distribution can be expressed in terms of the standard plus functions. Let us define the 
 dimensionless jet mass variable $\rho =M_J^2/Q^2$, where $Q = p_T^{J}R$. Then the following 
$\Lambda$-distribution with a given function $g(M_J^2)$ can be written as 
\be 
\Bigl[g(M_J^2)\Bigr]_{\Lambda^2}  = g(M_J^2) - \delta(M_J^2) \Bigl[\int^{\Lambda^2}_0 dM^2 g(M^2)\Bigr] 
= \frac{\hat{g}(\rho) }{Q^2} - \frac{\delta(\rho)}{Q^2} \Bigl[\int^{\Lambda^2/Q^2}_0 d\rho' g(\rho')\Bigr].
\ee 
Therefore if we define the dimensionless function, $\hat{g}(\rho) = Q^2 g(M_J^2)$, the distribution reads
\bea
Q^2\Bigl[g(M_J^2)\Bigr]_{\Lambda^2}  &=&  \hat{g}(\rho) - \delta(\rho) \Bigl[\int^1_0 d\rho' \hat{g}(\rho')
-\int^1_{\Lambda^2/Q^2} d\rho' \hat{g}(\rho')\Bigr] \nnb \\
&=&\Bigl[\hat{g}(\rho)\Bigr]_+ + \delta(\rho) \Bigl[\int^1_{\Lambda^2/Q^2} d\rho' \hat{g}(\rho')\Bigr].
\eea 
For example, the following $\Lambda$-distributions can be rewritten as  
\begin{eqnarray}
Q^2\Bigl[\frac{1}{M_J^2}\Bigr]_{\Lambda^2}  &=&  \frac{1}{\rho_+} -\delta (\rho) \ln \frac{\Lambda^2}{Q^2}, \nonumber \\
Q^2\Bigl[\frac{1}{M_J^2} \ln \frac{\mu^2}{M_J^2} \Bigr]_{\Lambda^2}  &=&   \frac{1}{\rho_+} \ln \frac{\mu^2}{Q^2} 
-\Bigl(\frac{\ln \rho}{\rho}\Bigr)_+ 
+\delta (\rho) \Bigl(\ln \frac{\Lambda^2}{Q^2} \ln \frac{\mu^2}{Q^2} -\frac{1}{2} \ln^2 \frac{\Lambda^2}{Q^2}\Bigr).
\end{eqnarray}

\section{NLO calculations of $\tilde{\mc{C}}_k$ and $\mc{S}_k^{II}$ with the 
zero-bin subtraction \label{apc}} 

\begin{figure*}[b]
\begin{center}
\includegraphics[width=15cm]{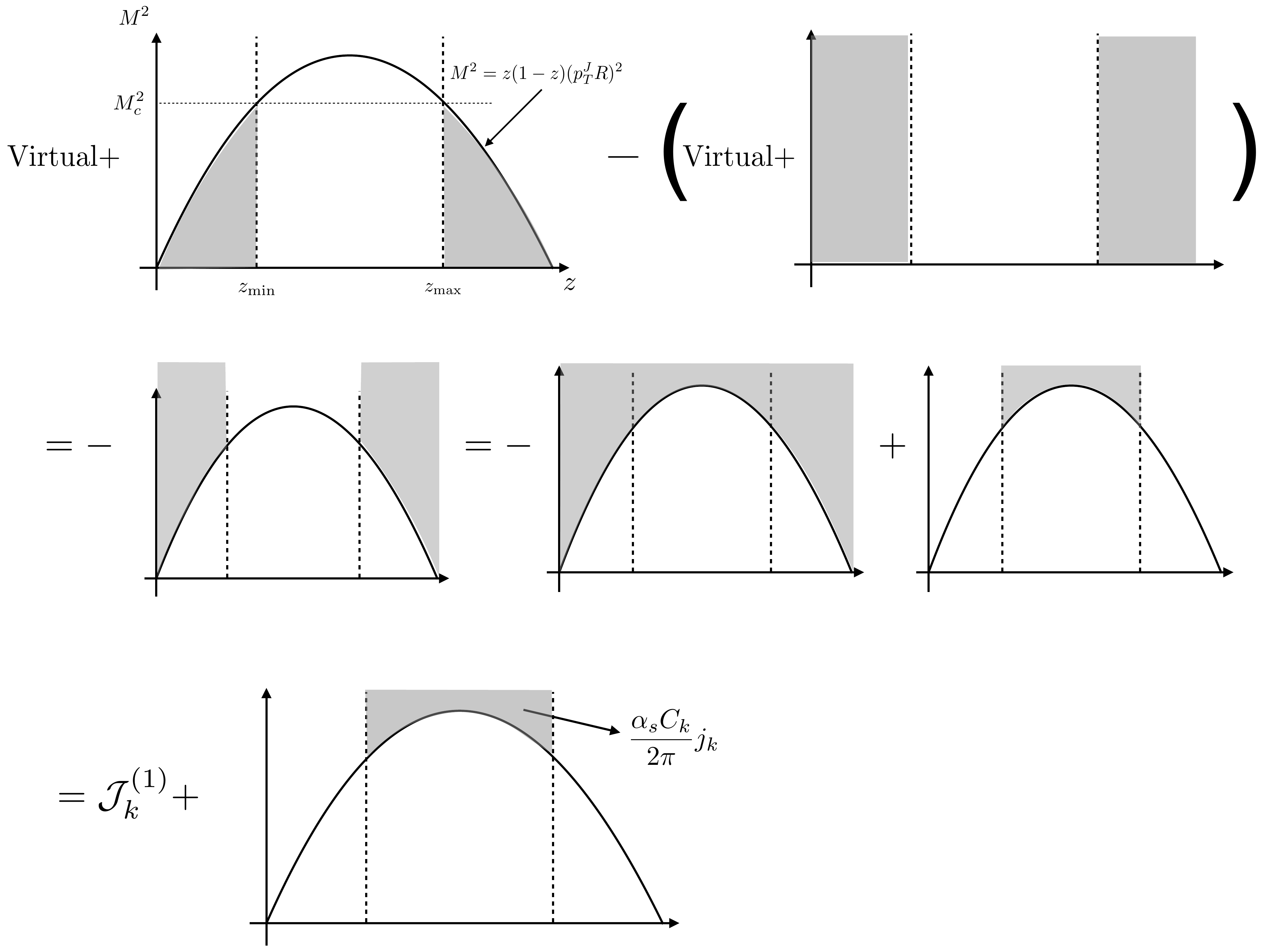}
\end{center}
\vspace{-0.6cm}
\caption{\label{fig6} \baselineskip 3.0ex 
The phase spaces for the collinear contribution in the region $\rho \ll y_c \sim \mathcal{O} (1)$ with the zero-bin subtraction. The resultant
phase space is shown in the second row and it is decomposed into two parts. The first one yields the integrated jet function since it is the sum
of the real contribution under the parabola with the virtual corrections. It is cancelled by performing the normalization, and 
the remaining second one yields the collinear function.
} 
\end{figure*}
 
The collinear contribution $\tilde{\mc{C}}_k (Q^2,\mu)$ in the limit $\rho \ll y_c \sim \mathcal{O} (1)$ can be computed by 
considering the phase spaces, as shown in Fig.~\ref{fig6}. The collinear emission cannot satisfy the mMDT criterion, otherwise $\rho$ 
becomes $\mO(1)$. Therefore it cannot contribute to the nonzero jet mass directly. 
In the first row in Fig.~\ref{fig6}, the figure on the left-hand side 
shows the possible phase space for the collinear emission along with the virtual corrections. 
The figure on the right-hand side is the zero-bin subtraction, in which the ultracollinear modes contribute. Since the ultracollinear modes do not 
recognize the boundary of the ungroomed jet (the parabola), the phase space extends to infinity for $M^2$. Since the virtual corrections in the collinear 
and the zero-bin contributions are the same,
they cancel through the subtraction and the available phase space is presented in the second row in Fig.~\ref{fig6}. 

The second figure in the second row yields
the integrated jet function since it corresponds to the real emission under the parabola with the virtual correction. Finally, with the normalization 
dividing by $\mc{J}_k$, the collinear
contribution with the zero-bin subtraction to NLO is given by 
\begin{equation}
\frac{1}{\mathcal{J}_k}  \Bigl(\mathcal{J}_k  + \frac{\alpha_s C_k}{2\pi} j_k\Bigr) = 1+\frac{\alpha_s C_k}{2\pi} j_k,
\end{equation}
where $j_k$ is given by
\begin{equation} \label{jk}
j_k (y_c, M_c^2) = \Bigl(\frac{1}{\eps} + \ln \frac{\mu^2}{M_c^2}\Bigr) g_k (y_c) + h_k (y_c) -I_k (y_c),
\end{equation}
and the $g_k,~h_k$ and $I_k$ are listed in Appendix~\ref{apa}.
From the structure of the phase space for $j_k$ in Fig.~\ref{fig6}, it is obvious that the pole in Eq.~\eqref{jk} is a UV divergence. Therefore  the collinear 
contribution $\tilde{\mc{C}}_k (y_c, Q^2,\mu)$ is given by Eq.~(\ref{hard2}) after removing the UV divergence in Eq.~(\ref{jk}).

\begin{figure*}[t]
\begin{center}
\includegraphics[width=16cm]{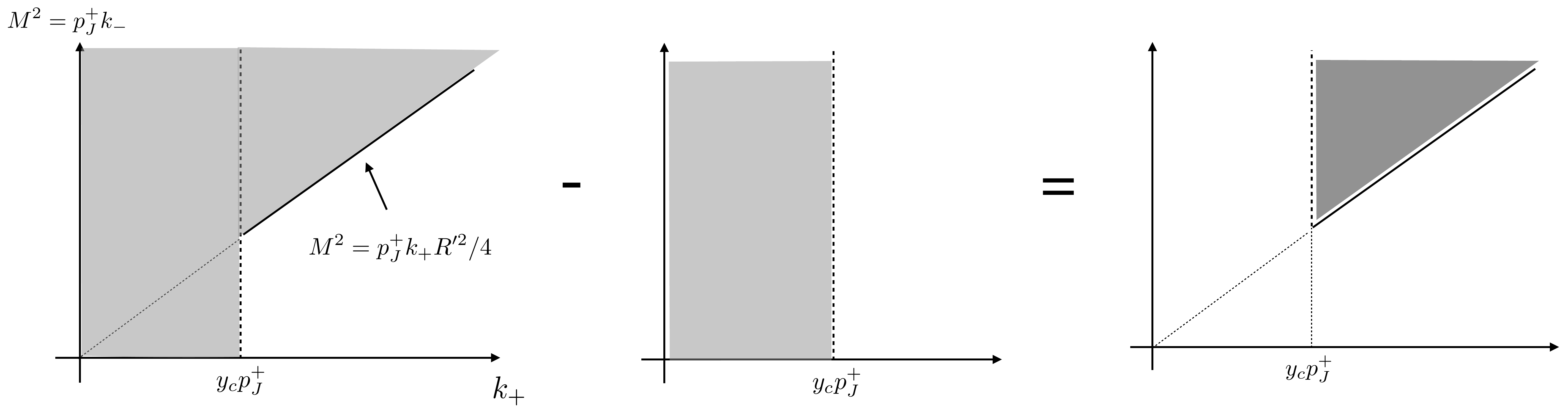}
\end{center}
\vspace{-0.6cm}
\caption{\label{fig7} \baselineskip 3.0ex 
The phases spaces for the csoft contributions with the zero-bin subtraction. The virtual contributions cancel since they are equal in the csoft
and its zero-bin contributions. The csoft function can be obtained by integrating over the shaded region of the phase space.
} 
\end{figure*}

The csoft function $\mc{S}_k^{II} (y_c^2Q^2,\mu)$ in the peak region ($\rho\ll y_c \ll 1$) can be computed by considering the phase spaces shown in Fig.~\ref{fig7}.
The first figure corresponds to the available phase space for the csoft gluon emission. Since the csoft radiation in this region cannot yield the nonzero groomed jet 
mass, it only contributes to the normalization of the jet mass distribution like $\tilde{\mc{C}}_k$. The second figure represents the phase space for the zero-bin 
subtraction. The virtual corrections in these two modes should be added, but they are the same. Therefore the virtual contributions cancel after the zero-bin subtraction. 
From the resultant phase space in Fig.~\ref{fig7}, we easily see that there is no IR divergence and only the UV divergence exists.  

The one loop result of the csoft contribution is obtained by integrating
over the shaded region in the third figure in Fig.~\ref{fig7}. The NLO results are given as
\begin{equation}
\mc{S}_k^{II} (y_c^2Q^2,\mu)=1+\frac{\alpha_s C_k}{2\pi}\Bigl(\frac{1}{\eps^2} +\frac{1}{\eps} \ln \frac{\mu^2}{y_c^2 Q^2} +\frac{1}{2} \ln^2 \frac{\mu^2}{y_c^2 Q^2}
-\frac{\pi^2}{12}\Bigr).
\end{equation}
After renormalization we obtain the csoft function $\mc{S}_k^{II} (y_c^2Q^2,\mu)$ in Eq.~(\ref{s2nlo}).

\section{NLO results of the factorized functions for the soft drop with $\beta > 0$ \label{apd}}

\begin{figure*}[b]
\begin{center}
\includegraphics[width=16cm]{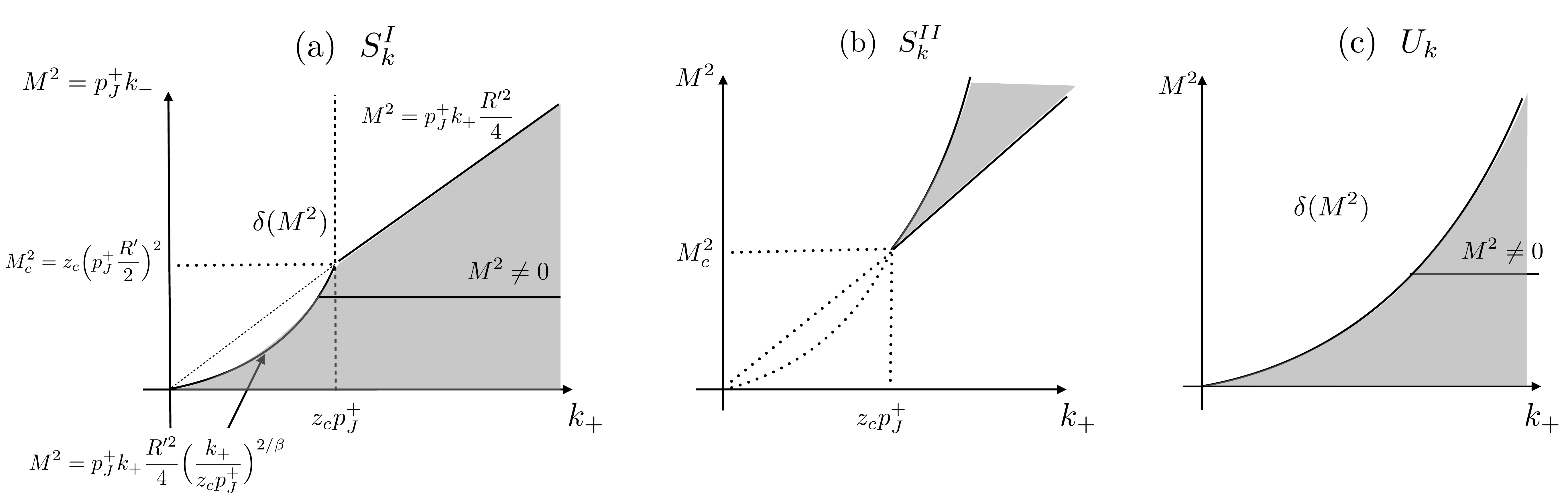} 
\end{center}
\vspace{-0.6cm}
\caption{\label{softdrop} 
The structure of the phase spaces  (a) $S_k^I (M^2)$ for the midrange region. (b) $\mc{S}_k^{II}$ and (c) $U_k (M^2)$ for the peak region. 
Here $z_c =\zc$. In (a) and (c), the phase spaces in the unshaded regions yield the results proportional to $\delta(M^2)$, while only the 
shaded region in (b) contributes. 
} 
\end{figure*}

We present the NLO results of the factorized functions for the groomed jet mass distributions using soft drop with $\beta> 0$ in the limit $\zc \ll 1$.
The soft drop condition is shown in Eq.~\eqref{sd}. 
Since the same factorization formulae as in mMDT can be applied to soft drop, Eq.~\eqref{factI} is the factorization theorem
for the midrange region $\rho \sim \zc \ll 1$ and Eq.~\eqref{factII} in the peak region $\rho \ll \zc \ll 1$. 
Since the functions $\mc{C}_k$ and $J_k$ are not affected by angular exponent $\beta$, 
we present $S_{k}^{I}$ in the midrange region and  $\mc{S}_k^{II}$ and $U_k$ in the peak region. 
The result in the midrange region is new. And the results for the peak region have been also computed in
Ref.~\cite{Kang:2018jwa}, but we could not compare the results directly since the exact definition of the plus distribution function employed in 
Ref.~\cite{Kang:2018jwa} was not mentioned.
 
\subsection{Midrange region: $\rho \sim z_c \ll 1$\label{sisd}}

The csoft function $S_k^I$ in the midrange region can be 
computed from the phase space illustrated in Fig.~\ref{softdrop}-(a).
For actual computation, we will consider the dimensionless csoft function $\bar{S}_k^{I} (\rho)$, where 
$\rho = M^2/Q^2$, and $Q = p_T^J R = p_J^+ R'/2$.
The part with $\delta(\rho)$ comes from the virtual contribution and the real contribution from the unshaded phase space in Fig.~\ref{softdrop}-(a).    st

And the part with $\rho\neq 0$ can be obtained by integrating $k_+$ over the shaded region in FIG.~\ref{softdrop}-(a). 
 When $\rho > \zc$, the contribution is given as
\begin{equation}
\label{mcs1}
M_{\cs} (\rho > \zc) = \frac{\alpha_s C_k}{\pi} \frac{1}{\rho} \Bigl( \frac{1}{\eps} +\ln \frac{\mu^2}{\rho^2 Q^2}\Bigr), 
\end{equation}
and when $\rho \le \zc$, it is written as
\begin{equation}
\label{mcs2}
M_{\cs} (\rho\leq \zc) = \frac{\alpha_s C_k}{\pi}\frac{e^{\gamma_E\eps}}{\Gamma(1-\eps)} 
\frac{1}{\eps}\Bigl(\frac{\mu^2}{Q^2}\Bigr)^{\eps} \zc^{-\frac{2\eps}{2+\beta}}
 \rho^{-1-\eps\frac{2+2\beta}{2+\beta}}. 
\end{equation}
Here the poles in Eqs.~\eqref{mcs1} and \eqref{mcs2} are of the UV origin. 
In Eq.~\eqref{mcs2}, there is an IR divergence as $\rho \to 0$, which is associated to the part with $\delta(\rho)$ 
after employing the standard plus distribution.  

Combining the results in Eqs.~\eqref{mcs1} and \eqref{mcs2} with the parts proportional to $\delta(\rho)$ (i.e, the virtual contribution and the real contribution 
from the unshaded region in Fig.~\ref{softdrop}-(a)), 
we obtain the (bare) NLO result of $\bar{S}_k^{I} (\rho)$ as 
\begin{align} 
\bar{S}_k^I (\rho,\mu)
&= \delta (\rho) +\frac{\alpha_s C_k}{2\pi} \Bigl\{ \delta (\rho) \Bigl(-\frac{1}{\eps^2} -\frac{1}{\eps}\ln\frac{\mu^2}{Q^2}-\frac{1}{2} \ln^2 \frac{\mu^2}{Q^2} +\frac{2}{2+\beta} \ln^2 z_c 
+\frac{\pi^2}{12}\Bigr)  \nonumber \\
\label{sdS1}
&+ \Bigl[ \frac{2}{\rho} \Bigl( \frac{1}{\eps}+\ln \frac{\mu^2}{\rho^2 Q^2} +\frac{2}{2+\beta} \ln \frac{\rho}{z_c}\Bigr)\Bigr]_+ 
-\frac{2}{2+\beta} \frac{2}{\rho} \ln \frac{\rho}{\zc} \Theta (\rho -\zc) \Bigr\}.
\end{align}
Here the IR divergence as $\rho \to 0$ in Eq.~\eqref{mcs2} is cancelled from the virtual correction, and all the remaining poles here are UV divergences. 

\subsection{Peak region: $\rho \ll z_c \ll 1$\label{apeak}}

The csoft contribution in the peak region can be obtained from the phase space, as shown in FIG.~\ref{softdrop}-(b). The phase space after the zero-bin subtraction is similar to the mMDT case (or the soft drop with $\beta=0$), but the vertical line ($k_+=y_c p_J^+$) in Fig.~\ref{fig7} is replaced by a
curve with $\beta>0$. From the phase space, 
it is obvious that the divergence is of the UV origin. Finally the bare csoft function $\bar{\mc{S}}_k^{II}$ is given to NLO as
\begin{equation}
\label{sdS2}
\bar{\mc{S}}_k^{II} (\mu) = 1+\frac{\alpha_s C_k}{2\pi} \frac{1}{1+\beta} \Bigl( 
\frac{1}{\eps^2}+\frac{1}{\eps} \ln \frac{\mu^2}{\zc^2 Q^2}+\frac{1}{2} \ln^2 \frac{\mu^2}{\zc^2 Q^2} 
-\frac{\pi^2}{12}\Bigr). 
\end{equation}

For the ucsoft function, the phase space is shown in Fig.~\ref{softdrop}-(c), in which the part with $\rho \neq 0$ comes from the shaded region,
while the rest of the phase space yields the $\delta (\rho)$ part. Following the method in computing $\bar{S}_k^I$, the dimensionless ucsoft functions
$\bar{U}_k (\rho)~(=Q^2 U_k(M^2))$ are given to NLO as 
\begin{eqnarray}
\bar{U}_k (\rho,\mu) &=& \delta (\rho) +\frac{\alpha_s C_k}{2\pi} \Bigl\{ -\frac{2+\beta}{1+\beta} \delta (\rho) \Bigl(\frac{1}{\eps^2} + \frac{1}{\eps} \ln \frac{\mu^2}{Q^2 \zc^{\frac{2}{2+\beta}}}+
\frac{1}{2} \ln^2 \frac{\mu^2}{Q^2 \zc^{\frac{2}{2+\beta}}} -\frac{\pi^2}{12}\Bigr) \nonumber \\
\label{sdU}
&&+ \Bigl[ \frac{2}{\rho} \Bigl(\frac{1}{\eps}+ \ln \frac{\mu^2}{Q^2 \zc^{\frac{2}{2+\beta}}} -\frac{2+2\beta}{2+\beta} \ln \rho\Bigr) \Bigr]_+\Bigr\}\ .
\end{eqnarray}
Here all the poles are UV divergences. 
From Eqs.~\eqref{sdS1}, \eqref{sdS2} and \eqref{sdU}, we easily check that  
the relation $\bar{S}_k^I = \bar{S}_k^{II} \bar{U}_k$ holds to NLO in $\as$. Therefore the results in the midrange and peak regions are consistent.

\begin{acknowledgments}
J. Chay is supported by Basic Science Research Program through the National Research Foundation of Korea (NRF) funded by 
the Ministry of Education (Grant No. NRF-2016R1D1A1B03935799). 
C.~Kim is supported by Basic Science Research Program through 
the National Research Foundation of Korea (NRF) funded by the Ministry of Science and ICT (Grant No. NRF-2017R1A2B4010511).
\end{acknowledgments}


\bibliographystyle{JHEP1}
\bibliography{Jet}


\end{document}